\documentclass[aps,prl,reprint,groupedaddress,showpacs,dvipdfmx]{revtex4-1} 
\usepackage{graphicx}
\usepackage{dcolumn}
\usepackage{bm}
\usepackage{amsmath}
\usepackage{comment}
\usepackage{afterpage}
\usepackage{amssymb}
\usepackage{txfonts}
\usepackage{url}
\usepackage{here}
\usepackage{multirow}
\usepackage[usenames,dvipsnames]{color}
\usepackage[version=3]{mhchem}
\definecolor{Gray}{gray}{0.0}
\definecolor{lightGray}{gray}{0.35}

\usepackage{chemfig}

\begin{document}
\title{Speeding up the {\it ab initio}
  diffusion Monte Carlo by a smart lattice regularization}

\author{Kousuke Nakano$^{1}$}
\email{kousuke\_1123@icloud.com}

\author{Ryo Maezono$^{2,3}$}

\author{Sandro Sorella$^{1}$}
\email{sorella@sissa.it}
  
\affiliation{$^{1}$ International School for Advanced Studies (SISSA),
  Via Bonomea 265, 34136, Trieste, Italy}

\affiliation{$^{2}$ School of Information Science, Japan Advanced Institute of Science and Technology (JAIST), Asahidai 1-1, Nomi, Ishikawa 923-1292, Japan}

\affiliation{$^{3}$ Computational Engineering Applications Unit, RIKEN, 2-1 Hirosawa, Wako, Saitama 351-0198, Japan}

\date{\today}

\begin{abstract}
One of the most significant drawbacks of the all-electron {\it ab initio} diffusion Monte Carlo (DMC) is that its computational cost drastically increases with the atomic number ($Z$), which typically scales with $Z^{\sim 6}$. In this study, we introduce an algorithm based on a very efficient implementation of the Lattice Regularized Diffusion Monte Carlo (LRDMC), where the conventional time discretization is replaced by its lattice space counterpart. This scheme enables us to conveniently adopt a small lattice space in the vicinity of nuclei, and a large one in the valence region, by which a considerable speedup is achieved, especially for large atomic number $Z$. Indeed, the computational performances 
of our algorithm can be theoretically established by using the Thomas-Fermi model for heavy atoms,  yielding an almost affordable scaling with the atomic number, {\it i.e.}, $Z^{\sim 5}$. This opens the way for efficient and accurate all-electron {\it ab initio} DMC in electronic structure calculations. 


\end{abstract}
\maketitle

\makeatletter
\def\Hline{%
\noalign{\ifnum0=`}\fi\hrule \@height 1pt \futurelet
\reserved@a\@xhline}
\makeatother

\makeatletter
\def\Bline{%
\noalign{\ifnum0=`}\fi\hrule \@height 0.5pt \futurelet
\reserved@a\@xhline}
\makeatother

%
%

%
\paragraph{Introduction $-$}
In recent years, the grand challenge in materials modeling
is to provide extremely accurate reference energetics
often well beyond the standard benchmark provided by the Density Functional Theory (DFT)
that notoriously is not enough predictive in several materials of
both scientific{\cite{2015KEI, 2017MAN}} and technological
interests~{\cite{2017NAU, 2018SOR}}.
This is also particularly important in view of existing progress in Machine Learning algorithms 
to define accurate classical force field potentials with reference data as unbiased as possible
\cite{2007BEH, 2017SCH, 2017YIN, 2017KOB}. 
For such problems, explicitly correlated wave-function-based
approaches are necessary~{\cite{1995ZHA, 2013BOO, 2016HOL, 2016HOL2, 2017CAR}},
such as the ones used in quantum chemistry and the ones relying
on statistical approaches that are known under the
generic name ``quantum Monte Carlo"
(QMC)~{\cite{2001FOU}}.
In practice, for electronic systems containing more
than a handful of atoms, QMC remains the only possible wave function based reference method, partly because of its favorable scaling with system size and the fact that it can be used efficiently on massively parallel supercomputers. One of the most powerful QMC techniques is based on a systematic ground state projection of a carefully determined trial state~{\cite{2017BEC}}, using the so-called diffusion Monte Carlo (DMC) with the fixed node approximation (FN). This choice represents a good compromise between accuracy and efficiency because FN is necessary for avoiding the well-known sign problem, and gives the best ({\it {i.e.}}, the lowest energy) variational state with the same sign of the trial function.
Despite this, FN remains a highly expensive computational tool,
especially for systems containing nuclei with large atomic number $Z$. 

\vspace{3mm}
In order to avoid an almost prohibitive computational cost, many sophisticated pseudopotentials for QMC calculations have been developed and intensively used so far{~\cite{2015TRA, 2016KRO, 2017TRA, 2017BEN, 2018BEN, 2018ANN}}. However, they are usually determined within other schemes and require further approximations ({\it e.g.}, the locality){\cite{{2001FOU}}} that spoil the consistency of the method and often sacrifices the variational principle.
At present, it is embarrassing to observe that several pseudopotentials used in QMC ({\it e.g.}, the so-called BFD ones~{\cite{2007BUR,2008BUR}}) are based on the Hartree-Fock (HF) approximation that completely misses the correlation energy. Their use can be therefore justified only empirically and does not guarantee any consistency, namely that the FN energy differences are consistent with or without pseudopotentials. 

\vspace{3mm}
All-electron calculations are rarely applied for atoms of large atomic
number $Z$ in QMC due to the expensive computational cost.
The major drawback of the all-electron calculations is that, 
in the electronic wave function, the core and the valence regions
are characterized by  very different length scales.
Therefore, within the most straightforward QMC algorithm,
the smallest scale ($\sim {Z^{ - 1}}$) should be adopted
for the proposed random displacement of the electrons,
in order to avoid significant biases.
Unfortunately, this implies several Markov iterations
to obtain a new uncorrelated sample, causing a high computational cost.
To solve this drawback, Umrigar {\it et al}.
have devised an accelerated Metropolis algorithm for the
variational Monte Carlo (VMC)~{\cite{1993UMR, 1998STE}},
by which electrons in the vicinity of nuclei are displaced
with a step much shorter than the one used in the valence region. 
They also developed another scheme for the diffusion Monte Carlo (DMC)~{\cite{1993UMR2}},
in which the velocity is reduced in the vicinity of nuclei to prevent
from overshooting  electrons.
Despite this improves the accuracy by a sizable amount, the major 
drawback of the conventional 
FN, is that the time step has to remain  necessarily 
the same both for the valence and the core region~{\cite{1993UMR2}}.
Instead, LRDMC can straightforwardly handle
different length scales of a wave function{\cite{2005CAS, 2017BEC}},
so that electrons in the vicinity of the nuclei and those in the
valence region can be appropriately diffused.
Henceforth this remedy is referred to as
``double-grid algorithm,"
as well as ``single-grid algorithm" refers
to the simpler version that adopts only a single lattice
space as introduced in Ref.~\onlinecite{2005CAS}.

\begin{figure}[htbp]
  \centering
  \includegraphics[width=7.6cm]{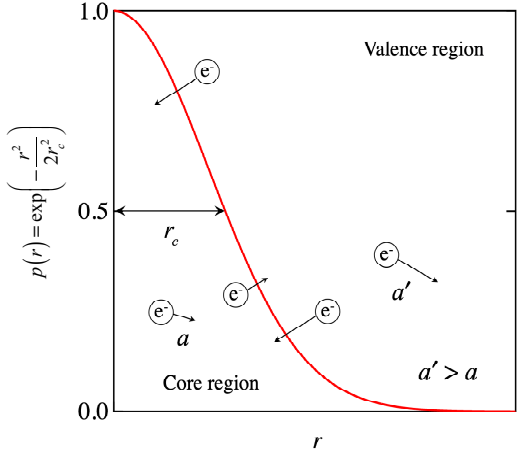}
  \caption{A two-dimensional schematic picture of the double-grid LRDMC algorithm. Each electron at ${\vec r}$ is displaced by a shorter ($a$) or a longer ($a'$) lattice space according to the probability of $p\left( {\vec r} \right)$ or $1 - p\left( {\vec r} \right)$, where $0 \leqslant p\left( {\vec r} \right) \leqslant 1$. Since $p\left( {\vec r} \right)$ is a function decaying suddenly far from nuclei ({\it e.g.}, Gaussian function), most electrons in the vicinity of nuclei are displaced by $a$, and those far from nuclei are by $a'$.}
  \label{double_grid_schematic}
\end{figure}

\vspace{3mm}
So far, the double-grid algorithm has been used only for a very
limited number of applications, specifically, for light elements
such as carbon~{\cite{2005CAS}} and sodium~{\cite{2019NAK}}.
For large atomic number ($Z$), too large computational
resources were required, also because the originally proposed algorithm
was very inefficient (see later).
In this study, we develop a generalized double-grid algorithm that drastically accelerates the calculation especially for large atomic number $Z$ without introducing biases, thus improving the computational scaling from $Z^{\sim 6}$ to $Z^{\sim 5}$.

%
%



\vspace{3mm}
\paragraph{Boosting the double-grid LRDMC $-$}
In LRDMC, the original continuous Hamiltonian is regularized
by an approximate one ${H^a}$ such that ${H^a} \to H$ for $a \to 0$,
where $a$ is the lattice mesh size used to discretize
the continuous space{~\cite{2005CAS, 2017BEC}}.
Indeed, the kinetic part is approximated by a finite difference form:
\begin{equation}
{\Delta _i} \approx \Delta _i^{a,a'} \equiv \Delta _i^{a,p} + \Delta _i^{a',1 - p},
\label{a_prime_a}
\end{equation}
where $\Delta _i^{a,p}$ and $\Delta _i^{a',1 - p}$ are discretized laplacians
by a small lattice space ($a$) and a large one ($a'$), respectively.
The function $p\left( {\vec r} \right)$, defining $\Delta _i^{a,p}$ and $\Delta _i^{a',1 - p}$,  
parametrizes the probability to use the smaller and 
therefore more accurate lattice space ($a$) when an electron is close to an heavy nucleus. 
In the previous works, $p(r)$ was chosen to be a simple Pade' function~{\cite{2005CAS}}:
\begin{equation}
p\left( {\vec r} \right) = {\left( {1 + {r_c}^2{{\left| {\vec r - {{\vec R}_c}} \right|}^2}} \right)^{ - 1}},
\label{p_poly}
\end{equation}
and a Gaussian-type function {\cite{2019NAK}}:
\begin{equation}
p\left( {\vec r} \right) = \exp \left( { - \frac{{{{\left| {\vec r - {{\vec R}_{\text{c}}}} \right|}^2}}}{{2{r_c}^2}}} \right)
\label{p_gaussian}
\end{equation}
where ${{\vec R}_\text{c}}$ is the position of the nucleus closest to the electron in $\vec r$, and $r_c$ is an important parameter determining the electrons treated with the smaller lattice space $a$ (henceforth referred to as core electrons), in other words, the ones inside  the sphere of radius $r_c$ (see Fig.~{\ref{double_grid_schematic}}).
This scheme enables us to use always the larger lattice space
$a'$ in the valence region, while the most expensive smaller
lattice space $a$ is used only when the electron is very close to the nucleus.

\vspace{3mm}
The key parameters of the double-grid LRDMC are $a'/a$ and $r_c$.
A smaller $r_c$ (larger $a'/a$) accelerates 
the double-grid scheme  
as compared with the corresponding single-grid one,
whereas the bias 
({\it i.e.}, the difference between
the single-grid and the double-grid LRDMC energies
at the same $a$) is correspondingly increased. 
Therefore, a proper determination of the two parameters is essential to balance accuracy and efficiency of the double-grid algorithm. $a'/a$ was originally parametrized as:
\begin{equation}
\frac{{a'}}{a} = \sqrt {\frac{{{Z^2}}}{4} + 1},
\label{a_prime_a_original}
\end{equation}
with the simple function $p\left( {\vec r} \right)$ in Eq.~(\ref{p_poly}) and $r_c$ = $\frac{1}{2}Z${~\cite{2005CAS}}, where $Z$ is an atomic number. However, as it is shown in the following, the above choice is not suitable for large atomic number ($Z$). 


\begin{center}
\begin{table*}[htbp]
\caption{\label{main_table_single_double_grid} LRDMC energies of He, Be, Ne, Ar, Kr and Xe atoms obtained by the single and double-grid schemes at $a = {\left( {3.5Z} \right)^{ - 1}}$.}
\vspace{3mm}
\begin{tabular}{c|c|c|c|c|c|c|c}
\Hline
\multirow{2}{*}{Element} & \multicolumn{2}{c|}{Lattice space} & \multicolumn{1}{c|}{Single grid} & \multicolumn{2}{c|}{Double grid (this work)} & \multicolumn{2}{c}{Double grid (previous)} \\
\cline{2-8}
 & $a \equiv \left( {\alpha  \cdot Z} \right)^{ - 1} $ &  $\alpha$ & Energy (Ha) & Energy (Ha) & Bias (mHa){\footnotemark[1]} & Energy (Ha) & Bias (mHa){\footnotemark[1]} \\
\Hline
He ($Z$ = 2) & 0.142857 & 3.50 & -2.9037321(62) & -2.9037434(63) & 0.0(0.0) & -2.9037398(63) & 0.0(0.0) \\
Be ($Z$ = 4) & 0.071429 & 3.50 & -14.667247(31) & -14.667330(32) & 0.1(0.0) & -14.667316(31) & 0.1(0.0) \\
Ne ($Z$ = 10) & 0.028571 & 3.50 & -128.92626(13) & -128.92647(14) & 0.2(0.2) & -128.92745(14) & 1.2(0.2) \\
Ar ($Z$ = 18) & 0.015873 & 3.50 & -527.49542(18) & -527.49686(19) & 1.4(0.3) & -527.50517(20) & 9.7(0.3) \\
Kr ($Z$ = 36) & 0.007937 & 3.50 & -2753.77151(78) & -2753.77133(70) & 0.2(1.0) & -2753.83069(84) & 59.2(1.1) \\
Xe ($Z$ = 54) & 0.005291 & 3.50 & -7234.8320(13) & -7234.8355(10) & 3.5(1.7) & -7235.0409(15) & 208.9(2.0) \\
\Hline
\end{tabular}

\footnotetext[1]{The difference in total energy between the single- and double-grid algorithms.}

\end{table*}
\end{center}

\begin{center}
\begin{table*}[htbp]
\caption{\label{table_benzene_single_double_grid} LRDMC energies of the benzene molecule (C$_6$H$_6$) obtained by the single and double-grid schemes at $a = {\left( {3.5Z_{\rm{max}}} \right)^{ - 1}}$.}
\vspace{3mm}
\begin{tabular}{c|c|c|c|c|c|c}
\Hline
\multirow{2}{*}{Molecule} & \multicolumn{2}{c|}{Lattice space} & \multicolumn{1}{c|}{Single grid} & \multicolumn{3}{c}{Double grid (this work)} \\
\cline{2-7}
 & $a \equiv \left( {\alpha  \cdot Z_{\rm{max}}} \right)^{ - 1} $ &  $\alpha$ & Energy (Ha) & Energy (Ha) & Bias (mHa){\footnotemark[1]} & Acceleration{\footnotemark[2]}\\
\Hline
C$_6$H$_6$ \fontsize{1.5pt}{0.0pt}\selectfont{\chemfig{[:0]**6(------)}} & 0.047619 & 3.50 & -232.19258(55) & -232.19424(56) & 1.7(8) & $\times$ 1.9 \\
\Hline
\end{tabular}

\footnotetext[1]{The difference in total energy between the single- and double-grid algorithms.}
\footnotetext[2]{The acceleration of actual CPU time required for a fixed reference error in the total energy.}

\end{table*}
\end{center}


\vspace{3mm}
In the following, we briefly describe the developed scheme.
First, we discuss a new strategy to properly determine $a'/a$. Since, the computational cost of LRDMC is proportional to the inverse square of the lattice spaces ($a$ and $a'$), the acceleration of the double-grid vs. single-grid LRDMC (denoted as speedup) can be analytically estimated in terms of $a^\prime/a$ and the average number of electrons in the core/valence regions, according to the following relation:
\begin{equation}
{\text{speedup}}^{ - 1} = \frac{{{N_{{\text{core}}}}\left( {{r_c}} \right)}}{Z} + \frac{{{N_{{\text{valence}}}}\left( {{r_c}} \right)}}{Z} \cdot {\left( {\frac{{a'}}{a}} \right)^{ - 2}},
\label{eq_acc_estimate}
\end{equation}
where ${{N_{{\rm{core}}}}\left( {{r_c}} \right)}$ and ${{N_{{\rm{valence}}}}\left( {{r_c}} \right)}=Z-{{N_{{\rm{core}}}}\left( {{r_c}} \right)}$ are the average numbers of electrons that are diffused with the smaller ($a$) and the larger ($a'$) lattice spaces, respectively
\footnote{The computational cost discussed here is not an actual CPU time but an acceptance ratio of off-diagonal trial moves in the Metropolis-Hastings algorithm. An actual CPU time is discussed later.}
.
On physical grounds, the average numbers of core and valence electrons satisfies the inequality ${N_{{\text{core}}}} \ll {N_{{\text{valence}}}}$.
On the other hand, the systematic error of the double-grid scheme referred to the corresponding single-grid one at the same $a$ (denoted as bias) cannot be analytically estimated. This is because it is a very complicated function of $a$, $a'$, $r_c$, and $Z$. It is, however, possible to estimate an appropriate value according to the following consideration:
Once $r_c$ and $a \simeq {1 \over Z}$ are given, it is clear that the corresponding bias increases with $a^\prime$. This implies that it is convenient to choose $a^\prime$ as small as possible as long as speedup$^{ - 1}$ does not sizably increase. 
Therefore, we determine $a^\prime/a$ in a way that the speedup becomes the half of the maximun ({\it i.e.}, ${\text{speedup}} = \frac{1}{2}{\left( {\frac{{{N_{{\text{core}}}}\left( {{r_c}} \right)}}{Z}} \right)^{ - 1}}$), yielding:
\begin{equation}
\frac{{a'}}{a} = \sqrt {\frac{{{N_{{\text{valence}}}}\left( {{r_c}} \right)}}{{{N_{{\text{core}}}}\left( {{r_c}} \right)}}}  \equiv \sqrt {\frac{{Z - {N_{{\text{core}}}}\left( {{r_c}} \right)}}{{{N_{{\text{core}}}}\left( {{r_c}} \right)}}}.
\label{a_prime_a_estimate}
\end{equation}
That the determination of $a'/a$ (Eq.~(\ref{a_prime_a_estimate})) corresponds to nearly the optimal compromise between efficiency and accuracy can be justified by the following argument: 
(i) If we chose a too large value of $a^\prime/a$, most of the computational time would be spent for the core electrons, and we could certainly decrease the bias by a smaller $a^\prime$ without affecting much the efficiency. ii) On  the other hand, if we chose $a^\prime/a$ ($>1$) too much close to one, the bias is minimal ({\it {i.e.}} equal to the single-grid algorithm), but the speedup can be substantially increased by a larger $a^\prime$,  a choice that should be clearly possible for the valence electrons.
Notice that, $N_{\rm{core}}\left( {{r_c}} \right)$ and $N_{\rm{valence}}\left( {{r_c}} \right)$ can be readily estimated from an atomic electron density calculated by an effective model such as the Thomas-Fermi~{\cite{1958LAN}} and the Slater ones~{\cite{1930SLA}}:
\begin{equation}
{N_{{\text{core}}}}\left( {{r_c}} \right) = \int_0^\infty  {dr\,\,4\pi {r^2}{\rho ^{{\text{eff}}}}\left( r \right)\exp \left( { - \frac{{{r^2}}}{{2{r_c}^2}}} \right)} .
\label{N_core}
\end{equation}
%


\vspace{3mm}
Next, we discuss a new strategy to properly determine $r_c$.
Since here we are interested in the  asymptotic behavior of the algorithmic accuracy and efficiency for large $Z$, it is convenient to adopt the Thomas-Fermi approximation~{\cite{1958LAN}}, according to which ${N_{{\text{core}}}}\left( {{r_c}} \right)$ is given by:
\begin{equation}
{N_{{\text{core}}}}\left( {{r_c}} \right) = \int_0^\infty  d r\,\,4\pi {r^2}{Z^2}f\left( {\frac{{{Z^{1/3}}r}}{b}} \right) \cdot \exp \left( { - \frac{{{r^2}}}{{2{r_c}^2}}} \right),
\end{equation}
where $b$ is a constant value and $f\left( x \right)$ is a universal function independent of $Z$. After integration (see the supplemental material in detail), we can obtain 
for $ {Z^{1/3}}{r_c}/b \ll 1$:
\begin{equation}
{N_{{\text{core}}}}\left( {{r_c}} \right) \propto {\left( {Z \cdot {r_c}} \right)^{\frac{3}{2}}}.
\label{nasympt}
\end{equation}
At this point, it is important to consider that the bias depends on the two lengths, namely, the value of $r_c$ (the bias is minimal for $r_c \to \infty$) and the value of $a^\prime$ (the bias is minimal if $a^\prime =a \simeq {1/Z}$).  Now, these two contributions are expected to be of the same order if we take $a^\prime \simeq r_c$ because we can assume that for $r>r_c$, far from the core region, the wave function is smooth and the laplacian can be discretized with a lattice space $ a^\prime \lesssim r_c$.  This represents the most balanced choice, providing a good compromise between efficiency (smaller $r_c$ and larger $a^\prime$) and accuracy (the other way around). 
With the above condition, by substituting the Thomas-Fermi expression of Eq.~(\ref{nasympt}) in Eq.~(\ref{a_prime_a_estimate}) for $a\propto {1/Z}$, we obtain:
\begin{equation}
{r_c} \gtrsim {a^\prime } \simeq \frac{1}{Z}\sqrt {\frac{Z}{{{{\left( {Z \cdot {r_c}} \right)}^{\frac{3}{2}}}}}},
\label{rc_a_scaling}
\end{equation}
yielding ${r_c} \propto {Z^{ - \theta }}$ with $\theta  = {5}/{7}$.
Therefore, our choice in the following 
is ${r_c}\left( Z \right) = \beta  \cdot {Z^{ - {5}/{7}}}$, where $\beta$ is a $Z$-independent prefactor.
Although the above discussion based on the Thomas-Fermi model is exact only for $Z \to \infty $, our VMC calculations show that the scaling ({\it {i.e.}}, ${N_{\rm{core}}}\left( {{r_c}} \right) \propto {Z^{3/7}}$) is undoubtedly correct even for small $Z$, as shown in Fig.~S-1 (see. supplemental material). The prefactor $\beta$ should be small enough so that the scaling is valid in a wide range of $Z$ values, even outside the asymptotic power law regime. Therefore, $\beta$ = 0.75 is employed in this study.


\vspace{3mm}
As a summary, in our algorithm, we determine ${r_c}\left( {Z} \right)$ according to ${r_c}\left( Z \right) = \beta \cdot {Z^{ - {5}/{7}}}$, with $\beta$ = 0.75. Then, a corresponding appropriate $a^\prime/a$ is determined  according to Eq.~({\ref{a_prime_a_estimate}}), wherein ${N_{\rm{core}}}\left( {{r_c}} \right)$ and ${N_{\rm{valence}}}\left( {{r_c}} \right)$ are estimated by the Slater's effective models~{\cite{1930SLA}} with the exponents that Clementi proposed, based on HF calculations~{\cite{1963CLE, 1967CLE}}. 
Since the computational cost of the all-electron single-grid DMC has turned out to scale with Z$^{5.5-6.5}$~{\cite{1986CEP, 1987HAM, 2005MA}}, and the single-grid LRDMC  similarly behaves, it is obviously very important to accelerate the double-grid LRDMC for heavy elements. In the following, we assume that the unbiased $ a \to 0 $ fixed node estimate can be obtained by a low order polynomial fit of several energy calculations corresponding to different $a\ge\simeq  { 1 \over Z}$. This implies, according to Eq.~(\ref{eq_acc_estimate}) and Eq.~({\ref{rc_a_scaling}}), that the new algorithm improves the complexity of the well known and widely used DMC algorithm by $\sim Z^{4/7} \simeq Z^{0.57}$, that  represents a remarkable achievement especially for large $Z$.

\vspace{3mm}
\paragraph{Practical test of the developed algorithm $-$}
In Table~{\ref{main_table_single_double_grid}}, we show the LRDMC energies of He, ($Z$ = 2), Be ($Z$ = 4), Ne ($Z$ = 10), Ar ($Z$ = 18), Kr ($Z$ = 36), and Xe ($Z$ = 54) atoms for $a = \left( {3.5 Z} \right)^{ - 1} $ obtained by the single-grid (standard), the previous and the newly developed double-grid algorithms. These results indicate that the double-grid LRDMC energies obtained with the previous parametrization are significantly biased, especially for large atomic number $Z$. On the other hand, our new parametrization suppresses these large biases, and the obtained LRDMC energies are essentially consistent with the single-grid ones for all $Z$, implying that the scaling law  derived 
by means of the Thomas-Fermi model (${r_c} \propto {Z^{ - 5/7}}$ and $a'/a \propto {Z^{2/7}}$) is in very good agreement with the numerical simulation. Thus, our newly developed double-grid algorithm accelerates the computation without introducing biases, no matter how large is $Z$. 

\begin{figure}[htbp]
  \centering
  \includegraphics[width=7.6cm]{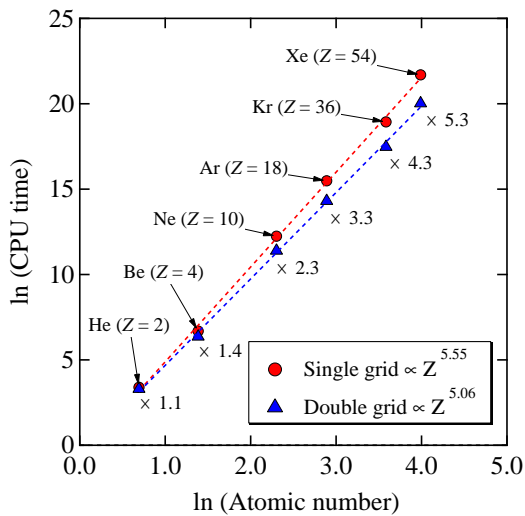}
  \caption{Scaling of the computational costs (CPU times) of the single and double-grid LRDMC, measured at $a$=${\left( {3.5Z} \right)^{ - 1}}$}
  \label{Practical-scaling}
\end{figure}

\vspace{3mm}
In practice, it is important to evaluate the actual computational time for a fixed reference error in the total energy, as a function of the atomic number Z. We measured the computational times for He, Be, Ne, Ar, Kr and Xe, wherein $a = {\left( {3.5Z} \right)^{ - 1}}$ is employed
{\footnote{
The calculations were performed on 8 Intel Xeon E5-2680v2-2.8~GHTz CPUs ({\it {i.e.}}, 320 cores) installed on a SGI cluster.
}}
.
This is consistent with the typical setting of the time step in the standard DMC ($\tau \propto {Z^{ - 2}}$~{\cite{2005MA}). Figure{~\ref{Practical-scaling}} shows that our new algorithm  accelerates the single-grid LRDMC calculations by $\times$ 1.1, $\times$ 1.4, $\times$ 2.3, $\times$ 3.3, $\times$ 4.3, and $\times$ 5.3 for He, Be, Ne, Ar, Kr, and Xe, respectively
\footnote{
The accelerations of actual CPU times are a slightly smaller than those of acceptance ratios ({\it {e.g.}}, $\times$ 3.9 and $\times$ 2.3 for CPU time and acceptance ratio, respectively, in the neon atom). 
This is because the double-grid algorithm consumes more CPU times when computing the discretized laplacians and potentials.
} 
.
Our practical test shows that the single-grid LRDMC scales with $Z^{5.55}$, which is already slightly better than the previous report for the standard DMC algorithm ($Z^{5.97}$ with $\tau \propto {Z^{ - 2}}$~{\cite{2005MA}}, where the Umrigar's improvement~{\cite{1993UMR2}} was employed), and the double-grid one improves the scaling to $Z^{5.06}$. The improvement of the computational time by the double-grid algorithm ($Z^{0.49}$) is consistent with our expectation ($Z^{4/7}$). To our best knowledge, $Z^{5.06}$ is the best scaling for the all-electron FN calculations so far.

%
\vspace{3mm}
\paragraph{Application to large systems $-$}
We discuss possible applications of the double-grid algorithm to large systems. For a polyatomic system, the smallest length scale is determined by the heaviest atom in the system with $Z=Z_{\rm max}$. Therefore, in this case, we can change the definition of $R_{c}$ in Eq.~({\ref{p_gaussian}}) slightly, by considering only the distances of the electrons with the heaviest atoms.
In this way, when electrons are close to the lighter elements, 
they  always move with the  larger lattice space $a'$, without introducing a 
sizable bias.
 Conversely, for $r_c$, one can adopt the value calculated with a single reference heavy atom, as we have done in this work. It is clear, therefore, that a  more significant speedup can be achieved by using Eq~({\ref{eq_acc_estimate}}), especially when the number of heavy atoms in the system is very small ({\it {e.g.}}, transition-metal porphyrin complexes, metallofullerenes). As the first step to large systems, we considered the benzene molecule ($Z_{\rm{max}}$ = 6). Table~{\ref{table_benzene_single_double_grid}} shows that the bias of the double-grid LRDMC is as small as in the atomic cases while the computational time is accelerated by $\times$ 1.9, significantly larger than the one $\simeq 1.5$ estimated from Fig.~\ref{Practical-scaling}}, demonstrating that the double-grid algorithm is already  
 advantageous for polyatomic systems, even without too heavy nuclei and too many light ones.
Finally, we have compared the computational costs between all-electron and pseudopotential calculations
\footnote{
For the benzene molecule, we measured the CPU times for a fixed reference error in the total energy at $a$ = 0.1 Bohr and $a$ = 0.3 Bohr for the all-electron and the pseudopotential calculations, respectively. We determined these values ($a_{\rm{min}}$) such that the extrapolation error obtained with 3 parameter polynomial fit ($E\left( a \right) = {E_0} + {k_1} \cdot {a^2} + {k_2} \cdot {a^4}$) of independent energy calculations corresponding to 8 different values of $a \ge a_{\rm{min}}$ becomes $\sim$ 2.0 mHa referenced to the safest extrapolation value ({\it {i.e.}}, the smallest $a_{\rm{min}}$). Notice that the energy-consistent BFD pseudopotentials with the VDZ basis were employed.
}
.
Thanks to the significant acceleration, the CPU time of the all-electron double-grid LRDMC for the benzene molecule is just 5.7 times larger than the pseudo-potential single-grid one. Thus, the double-grid LRDMC should make possible the application of the all-electron FN to realistic materials, allowing extremely accurate and easily reproducible reference energies in the future.

\vspace{3mm}
\paragraph{Summary $-$}
In this study, we develop a new generalized algorithm of the double-grid Lattice Regularized Diffusion Monte Carlo (LRDMC). The speedup of the algorithm is predicted theoretically within the standard Thomas-Fermi model for atoms with large atomic number, and the calculation is indeed accelerated in practice  by a large amount, especially for large atomic number $Z$ and without inducing significant biases. As a result, the computational scaling is improved from $Z^{5.55}$ to $Z^{5.06}$. Our double-grid algorithm can be applied to polyatomic systems with further significant speedups. Last but not least it should be possible with the present technique to treat ions and electrons without relying on the Born-Oppenheimer approximation, because the corresponding much different length scales should be efficiently considered within the proposed double-grid scheme.

\begin{acknowledgments}
\vspace{3mm}
\paragraph{Acknowledgments $-$}
The computations in this work have been mainly performed using the facilities of Research Center for Advanced Computing Infrastructure at Japan Advanced Institute of Science and Technology (JAIST). K. Nakano is grateful for a financial support from Simons Foundation. R. Maezono is grateful for financial supports from MEXT-KAKENHI (19H04692 and 16KK0097), from FLAGSHIP2020 (project nos. hp190169 and hp190167 at K-computer), from Toyota Motor Corporation, from I-O DATA Foundation, from the Air Force Office of Scientific Research 
(AFOSR-AOARD/FA2386-17-1-4049;FA2386-19-1-4015), and from JSPS Bilateral Joint Projects (with India DST).
\end{acknowledgments}

\bibliographystyle{apsrev4-1}
\bibliography{references.bib}

\begin{thebibliography}{39}%
\makeatletter
\providecommand \@ifxundefined [1]{%
 \@ifx{#1\undefined}
}%
\providecommand \@ifnum [1]{%
 \ifnum #1\expandafter \@firstoftwo
 \else \expandafter \@secondoftwo
 \fi
}%
\providecommand \@ifx [1]{%
 \ifx #1\expandafter \@firstoftwo
 \else \expandafter \@secondoftwo
 \fi
}%
\providecommand \natexlab [1]{#1}%
\providecommand \enquote  [1]{``#1''}%
\providecommand \bibnamefont  [1]{#1}%
\providecommand \bibfnamefont [1]{#1}%
\providecommand \citenamefont [1]{#1}%
\providecommand \href@noop [0]{\@secondoftwo}%
\providecommand \href [0]{\begingroup \@sanitize@url \@href}%
\providecommand \@href[1]{\@@startlink{#1}\@@href}%
\providecommand \@@href[1]{\endgroup#1\@@endlink}%
\providecommand \@sanitize@url [0]{\catcode `\\12\catcode `\$12\catcode
  `\&12\catcode `\#12\catcode `\^12\catcode `\_12\catcode `\%12\relax}%
\providecommand \@@startlink[1]{}%
\providecommand \@@endlink[0]{}%
\providecommand \url  [0]{\begingroup\@sanitize@url \@url }%
\providecommand \@url [1]{\endgroup\@href {#1}{\urlprefix }}%
\providecommand \urlprefix  [0]{URL }%
\providecommand \Eprint [0]{\href }%
\providecommand \doibase [0]{http://dx.doi.org/}%
\providecommand \selectlanguage [0]{\@gobble}%
\providecommand \bibinfo  [0]{\@secondoftwo}%
\providecommand \bibfield  [0]{\@secondoftwo}%
\providecommand \translation [1]{[#1]}%
\providecommand \BibitemOpen [0]{}%
\providecommand \bibitemStop [0]{}%
\providecommand \bibitemNoStop [0]{.\EOS\space}%
\providecommand \EOS [0]{\spacefactor3000\relax}%
\providecommand \BibitemShut  [1]{\csname bibitem#1\endcsname}%
\let\auto@bib@innerbib\@empty
\bibitem [{\citenamefont {Keimer}\ \emph {et~al.}(2015)\citenamefont {Keimer},
  \citenamefont {Kivelson}, \citenamefont {Norman}, \citenamefont {Uchida},\
  and\ \citenamefont {Zaanen}}]{2015KEI}%
  \BibitemOpen
  \bibfield  {author} {\bibinfo {author} {\bibfnamefont {B.}~\bibnamefont
  {Keimer}}, \bibinfo {author} {\bibfnamefont {S.~A.}\ \bibnamefont
  {Kivelson}}, \bibinfo {author} {\bibfnamefont {M.~R.}\ \bibnamefont
  {Norman}}, \bibinfo {author} {\bibfnamefont {S.}~\bibnamefont {Uchida}}, \
  and\ \bibinfo {author} {\bibfnamefont {J.}~\bibnamefont {Zaanen}},\ }\href
  {\doibase 10.1038/nature14165} {\bibfield  {journal} {\bibinfo  {journal}
  {Nature}\ }\textbf {\bibinfo {volume} {518}},\ \bibinfo {pages} {179}
  (\bibinfo {year} {2015})}\BibitemShut {NoStop}%
\bibitem [{\citenamefont {Manzeli}\ \emph {et~al.}(2017)\citenamefont
  {Manzeli}, \citenamefont {Ovchinnikov}, \citenamefont {Pasquier},
  \citenamefont {Yazyev},\ and\ \citenamefont {Kis}}]{2017MAN}%
  \BibitemOpen
  \bibfield  {author} {\bibinfo {author} {\bibfnamefont {S.}~\bibnamefont
  {Manzeli}}, \bibinfo {author} {\bibfnamefont {D.}~\bibnamefont
  {Ovchinnikov}}, \bibinfo {author} {\bibfnamefont {D.}~\bibnamefont
  {Pasquier}}, \bibinfo {author} {\bibfnamefont {O.~V.}\ \bibnamefont
  {Yazyev}}, \ and\ \bibinfo {author} {\bibfnamefont {A.}~\bibnamefont {Kis}},\
  }\href {\doibase 10.1038/natrevmats.2017.33} {\bibfield  {journal} {\bibinfo
  {journal} {Nat. Rev. Mater.}\ }\textbf {\bibinfo {volume} {2}},\ \bibinfo
  {pages} {17033} (\bibinfo {year} {2017})}\BibitemShut {NoStop}%
\bibitem [{\citenamefont {Naumis}\ \emph {et~al.}(2017)\citenamefont {Naumis},
  \citenamefont {Barraza-Lopez}, \citenamefont {Oliva-Leyva},\ and\
  \citenamefont {Terrones}}]{2017NAU}%
  \BibitemOpen
  \bibfield  {author} {\bibinfo {author} {\bibfnamefont {G.~G.}\ \bibnamefont
  {Naumis}}, \bibinfo {author} {\bibfnamefont {S.}~\bibnamefont
  {Barraza-Lopez}}, \bibinfo {author} {\bibfnamefont {M.}~\bibnamefont
  {Oliva-Leyva}}, \ and\ \bibinfo {author} {\bibfnamefont {H.}~\bibnamefont
  {Terrones}},\ }\href {\doibase 10.1088/1361-6633/aa74ef} {\bibfield
  {journal} {\bibinfo  {journal} {Reports Prog. Phys.}\ }\textbf {\bibinfo
  {volume} {80}},\ \bibinfo {pages} {096501} (\bibinfo {year}
  {2017})}\BibitemShut {NoStop}%
\bibitem [{\citenamefont {Sorella}\ \emph {et~al.}(2018)\citenamefont
  {Sorella}, \citenamefont {Seki}, \citenamefont {Brovko}, \citenamefont
  {Shirakawa}, \citenamefont {Miyakoshi}, \citenamefont {Yunoki},\ and\
  \citenamefont {Tosatti}}]{2018SOR}%
  \BibitemOpen
  \bibfield  {author} {\bibinfo {author} {\bibfnamefont {S.}~\bibnamefont
  {Sorella}}, \bibinfo {author} {\bibfnamefont {K.}~\bibnamefont {Seki}},
  \bibinfo {author} {\bibfnamefont {O.~O.}\ \bibnamefont {Brovko}}, \bibinfo
  {author} {\bibfnamefont {T.}~\bibnamefont {Shirakawa}}, \bibinfo {author}
  {\bibfnamefont {S.}~\bibnamefont {Miyakoshi}}, \bibinfo {author}
  {\bibfnamefont {S.}~\bibnamefont {Yunoki}}, \ and\ \bibinfo {author}
  {\bibfnamefont {E.}~\bibnamefont {Tosatti}},\ }\href {\doibase
  10.1103/PhysRevLett.121.066402} {\bibfield  {journal} {\bibinfo  {journal}
  {Phys. Rev. Lett.}\ }\textbf {\bibinfo {volume} {121}},\ \bibinfo {pages}
  {066402} (\bibinfo {year} {2018})}\BibitemShut {NoStop}%
\bibitem [{\citenamefont {Behler}\ and\ \citenamefont
  {Parrinello}(2007)}]{2007BEH}%
  \BibitemOpen
  \bibfield  {author} {\bibinfo {author} {\bibfnamefont {J.}~\bibnamefont
  {Behler}}\ and\ \bibinfo {author} {\bibfnamefont {M.}~\bibnamefont
  {Parrinello}},\ }\href {\doibase 10.1103/PhysRevLett.98.146401} {\bibfield
  {journal} {\bibinfo  {journal} {Phys. Rev. Lett.}\ }\textbf {\bibinfo
  {volume} {98}},\ \bibinfo {pages} {146401} (\bibinfo {year}
  {2007})}\BibitemShut {NoStop}%
\bibitem [{\citenamefont {Schmidt}\ \emph {et~al.}(2017)\citenamefont
  {Schmidt}, \citenamefont {Shi}, \citenamefont {Borlido}, \citenamefont
  {Chen}, \citenamefont {Botti},\ and\ \citenamefont {Marques}}]{2017SCH}%
  \BibitemOpen
  \bibfield  {author} {\bibinfo {author} {\bibfnamefont {J.}~\bibnamefont
  {Schmidt}}, \bibinfo {author} {\bibfnamefont {J.}~\bibnamefont {Shi}},
  \bibinfo {author} {\bibfnamefont {P.}~\bibnamefont {Borlido}}, \bibinfo
  {author} {\bibfnamefont {L.}~\bibnamefont {Chen}}, \bibinfo {author}
  {\bibfnamefont {S.}~\bibnamefont {Botti}}, \ and\ \bibinfo {author}
  {\bibfnamefont {M.~A.}\ \bibnamefont {Marques}},\ }\href@noop {} {\bibfield
  {journal} {\bibinfo  {journal} {Chem. Mater.}\ }\textbf {\bibinfo {volume}
  {29}},\ \bibinfo {pages} {5090} (\bibinfo {year} {2017})}\BibitemShut
  {NoStop}%
\bibitem [{\citenamefont {Li}\ \emph {et~al.}(2017)\citenamefont {Li},
  \citenamefont {Li}, \citenamefont {Pickard~IV}, \citenamefont {Narayanan},
  \citenamefont {Sen}, \citenamefont {Chan}, \citenamefont {Sankaranarayanan},
  \citenamefont {Brooks},\ and\ \citenamefont {Roux}}]{2017YIN}%
  \BibitemOpen
  \bibfield  {author} {\bibinfo {author} {\bibfnamefont {Y.}~\bibnamefont
  {Li}}, \bibinfo {author} {\bibfnamefont {H.}~\bibnamefont {Li}}, \bibinfo
  {author} {\bibfnamefont {F.~C.}\ \bibnamefont {Pickard~IV}}, \bibinfo
  {author} {\bibfnamefont {B.}~\bibnamefont {Narayanan}}, \bibinfo {author}
  {\bibfnamefont {F.~G.}\ \bibnamefont {Sen}}, \bibinfo {author} {\bibfnamefont
  {M.~K.}\ \bibnamefont {Chan}}, \bibinfo {author} {\bibfnamefont {S.~K.}\
  \bibnamefont {Sankaranarayanan}}, \bibinfo {author} {\bibfnamefont {B.~R.}\
  \bibnamefont {Brooks}}, \ and\ \bibinfo {author} {\bibfnamefont
  {B.}~\bibnamefont {Roux}},\ }\href@noop {} {\bibfield  {journal} {\bibinfo
  {journal} {J. Chem. Theory Comput.}\ }\textbf {\bibinfo {volume} {13}},\
  \bibinfo {pages} {4492} (\bibinfo {year} {2017})}\BibitemShut {NoStop}%
\bibitem [{\citenamefont {Kobayashi}\ \emph {et~al.}(2017)\citenamefont
  {Kobayashi}, \citenamefont {Giofr\'e}, \citenamefont {Junge}, \citenamefont
  {Ceriotti},\ and\ \citenamefont {Curtin}}]{2017KOB}%
  \BibitemOpen
  \bibfield  {author} {\bibinfo {author} {\bibfnamefont {R.}~\bibnamefont
  {Kobayashi}}, \bibinfo {author} {\bibfnamefont {D.}~\bibnamefont {Giofr\'e}},
  \bibinfo {author} {\bibfnamefont {T.}~\bibnamefont {Junge}}, \bibinfo
  {author} {\bibfnamefont {M.}~\bibnamefont {Ceriotti}}, \ and\ \bibinfo
  {author} {\bibfnamefont {W.~A.}\ \bibnamefont {Curtin}},\ }\href {\doibase
  10.1103/PhysRevMaterials.1.053604} {\bibfield  {journal} {\bibinfo  {journal}
  {Phys. Rev. Materials}\ }\textbf {\bibinfo {volume} {1}},\ \bibinfo {pages}
  {053604} (\bibinfo {year} {2017})}\BibitemShut {NoStop}%
\bibitem [{\citenamefont {Zhang}\ \emph {et~al.}(1995)\citenamefont {Zhang},
  \citenamefont {Carlson},\ and\ \citenamefont {Gubernatis}}]{1995ZHA}%
  \BibitemOpen
  \bibfield  {author} {\bibinfo {author} {\bibfnamefont {S.}~\bibnamefont
  {Zhang}}, \bibinfo {author} {\bibfnamefont {J.}~\bibnamefont {Carlson}}, \
  and\ \bibinfo {author} {\bibfnamefont {J.~E.}\ \bibnamefont {Gubernatis}},\
  }\href {https://journals.aps.org/prl/pdf/10.1103/PhysRevLett.74.3652}
  {\bibfield  {journal} {\bibinfo  {journal} {Phys. Rev. Lett.}\ }\textbf
  {\bibinfo {volume} {74}},\ \bibinfo {pages} {3652} (\bibinfo {year}
  {1995})}\BibitemShut {NoStop}%
\bibitem [{\citenamefont {Booth}\ \emph {et~al.}(2013)\citenamefont {Booth},
  \citenamefont {Gr{\"{u}}neis}, \citenamefont {Kresse},\ and\ \citenamefont
  {Alavi}}]{2013BOO}%
  \BibitemOpen
  \bibfield  {author} {\bibinfo {author} {\bibfnamefont {G.~H.}\ \bibnamefont
  {Booth}}, \bibinfo {author} {\bibfnamefont {A.}~\bibnamefont
  {Gr{\"{u}}neis}}, \bibinfo {author} {\bibfnamefont {G.}~\bibnamefont
  {Kresse}}, \ and\ \bibinfo {author} {\bibfnamefont {A.}~\bibnamefont
  {Alavi}},\ }\href {\doibase 10.1038/nature11770} {\bibfield  {journal}
  {\bibinfo  {journal} {Nature}\ }\textbf {\bibinfo {volume} {493}},\ \bibinfo
  {pages} {365} (\bibinfo {year} {2013})}\BibitemShut {NoStop}%
\bibitem [{\citenamefont {Holmes}\ \emph
  {et~al.}(2016{\natexlab{a}})\citenamefont {Holmes}, \citenamefont
  {Changlani},\ and\ \citenamefont {Umrigar}}]{2016HOL}%
  \BibitemOpen
  \bibfield  {author} {\bibinfo {author} {\bibfnamefont {A.~A.}\ \bibnamefont
  {Holmes}}, \bibinfo {author} {\bibfnamefont {H.~J.}\ \bibnamefont
  {Changlani}}, \ and\ \bibinfo {author} {\bibfnamefont {C.~J.}\ \bibnamefont
  {Umrigar}},\ }\href {\doibase 10.1021/acs.jctc.5b01170} {\bibfield  {journal}
  {\bibinfo  {journal} {J. Chem. Theory Comput.}\ }\textbf {\bibinfo {volume}
  {12}},\ \bibinfo {pages} {1561} (\bibinfo {year}
  {2016}{\natexlab{a}})}\BibitemShut {NoStop}%
\bibitem [{\citenamefont {Holmes}\ \emph
  {et~al.}(2016{\natexlab{b}})\citenamefont {Holmes}, \citenamefont {Tubman},\
  and\ \citenamefont {Umrigar}}]{2016HOL2}%
  \BibitemOpen
  \bibfield  {author} {\bibinfo {author} {\bibfnamefont {A.~A.}\ \bibnamefont
  {Holmes}}, \bibinfo {author} {\bibfnamefont {N.~M.}\ \bibnamefont {Tubman}},
  \ and\ \bibinfo {author} {\bibfnamefont {C.~J.}\ \bibnamefont {Umrigar}},\
  }\href {\doibase 10.1021/acs.jctc.6b00407} {\bibfield  {journal} {\bibinfo
  {journal} {J. Chem. Theory Comput.}\ }\textbf {\bibinfo {volume} {12}},\
  \bibinfo {pages} {3674} (\bibinfo {year} {2016}{\natexlab{b}})}\BibitemShut
  {NoStop}%
\bibitem [{\citenamefont {Carleo}\ and\ \citenamefont
  {Troyer}(2017)}]{2017CAR}%
  \BibitemOpen
  \bibfield  {author} {\bibinfo {author} {\bibfnamefont {G.}~\bibnamefont
  {Carleo}}\ and\ \bibinfo {author} {\bibfnamefont {M.}~\bibnamefont
  {Troyer}},\ }\href {http://science.sciencemag.org/} {\bibfield  {journal}
  {\bibinfo  {journal} {Science}\ }\textbf {\bibinfo {volume} {355}},\ \bibinfo
  {pages} {602} (\bibinfo {year} {2017})}\BibitemShut {NoStop}%
\bibitem [{\citenamefont {Foulkes}\ \emph {et~al.}(2001)\citenamefont
  {Foulkes}, \citenamefont {Mitas}, \citenamefont {Needs},\ and\ \citenamefont
  {Rajagopal}}]{2001FOU}%
  \BibitemOpen
  \bibfield  {author} {\bibinfo {author} {\bibfnamefont {W.}~\bibnamefont
  {Foulkes}}, \bibinfo {author} {\bibfnamefont {L.}~\bibnamefont {Mitas}},
  \bibinfo {author} {\bibfnamefont {R.}~\bibnamefont {Needs}}, \ and\ \bibinfo
  {author} {\bibfnamefont {G.}~\bibnamefont {Rajagopal}},\ }\href@noop {}
  {\bibfield  {journal} {\bibinfo  {journal} {Rev. Mod. Phys.}\ }\textbf
  {\bibinfo {volume} {73}},\ \bibinfo {pages} {33} (\bibinfo {year}
  {2001})}\BibitemShut {NoStop}%
\bibitem [{\citenamefont {Becca}\ and\ \citenamefont
  {Sorella}(2017)}]{2017BEC}%
  \BibitemOpen
  \bibfield  {author} {\bibinfo {author} {\bibfnamefont {F.}~\bibnamefont
  {Becca}}\ and\ \bibinfo {author} {\bibfnamefont {S.}~\bibnamefont
  {Sorella}},\ }\href@noop {} {\emph {\bibinfo {title} {{Quantum Monte Carlo
  approaches for correlated systems}}}}\ (\bibinfo  {publisher} {Cambridge
  University Press},\ \bibinfo {year} {2017})\BibitemShut {NoStop}%
\bibitem [{\citenamefont {Trail}\ and\ \citenamefont {Needs}(2015)}]{2015TRA}%
  \BibitemOpen
  \bibfield  {author} {\bibinfo {author} {\bibfnamefont {J.~R.}\ \bibnamefont
  {Trail}}\ and\ \bibinfo {author} {\bibfnamefont {R.~J.}\ \bibnamefont
  {Needs}},\ }\href {\doibase 10.1063/1.4907589} {\bibfield  {journal}
  {\bibinfo  {journal} {J. Chem. Phys.}\ }\textbf {\bibinfo {volume} {142}},\
  \bibinfo {pages} {064110} (\bibinfo {year} {2015})}\BibitemShut {NoStop}%
\bibitem [{\citenamefont {Krogel}\ \emph {et~al.}(2016)\citenamefont {Krogel},
  \citenamefont {Santana},\ and\ \citenamefont {Reboredo}}]{2016KRO}%
  \BibitemOpen
  \bibfield  {author} {\bibinfo {author} {\bibfnamefont {J.~T.}\ \bibnamefont
  {Krogel}}, \bibinfo {author} {\bibfnamefont {J.~A.}\ \bibnamefont {Santana}},
  \ and\ \bibinfo {author} {\bibfnamefont {F.~A.}\ \bibnamefont {Reboredo}},\
  }\href {\doibase 10.1103/PhysRevB.93.075143} {\bibfield  {journal} {\bibinfo
  {journal} {Phys. Rev. B}\ }\textbf {\bibinfo {volume} {93}},\ \bibinfo
  {pages} {75143} (\bibinfo {year} {2016})}\BibitemShut {NoStop}%
\bibitem [{\citenamefont {Trail}\ and\ \citenamefont {Needs}(2017)}]{2017TRA}%
  \BibitemOpen
  \bibfield  {author} {\bibinfo {author} {\bibfnamefont {J.~R.}\ \bibnamefont
  {Trail}}\ and\ \bibinfo {author} {\bibfnamefont {R.~J.}\ \bibnamefont
  {Needs}},\ }\href {\doibase 10.1063/1.4984046} {\bibfield  {journal}
  {\bibinfo  {journal} {J. Chem. Phys.}\ }\textbf {\bibinfo {volume} {146}},\
  \bibinfo {pages} {204107} (\bibinfo {year} {2017})}\BibitemShut {NoStop}%
\bibitem [{\citenamefont {Bennett}\ \emph {et~al.}(2017)\citenamefont
  {Bennett}, \citenamefont {Melton}, \citenamefont {Annaberdiyev},
  \citenamefont {Wang}, \citenamefont {Shulenburger},\ and\ \citenamefont
  {Mitas}}]{2017BEN}%
  \BibitemOpen
  \bibfield  {author} {\bibinfo {author} {\bibfnamefont {M.~C.}\ \bibnamefont
  {Bennett}}, \bibinfo {author} {\bibfnamefont {C.~A.}\ \bibnamefont {Melton}},
  \bibinfo {author} {\bibfnamefont {A.}~\bibnamefont {Annaberdiyev}}, \bibinfo
  {author} {\bibfnamefont {G.}~\bibnamefont {Wang}}, \bibinfo {author}
  {\bibfnamefont {L.}~\bibnamefont {Shulenburger}}, \ and\ \bibinfo {author}
  {\bibfnamefont {L.}~\bibnamefont {Mitas}},\ }\href {\doibase
  10.1063/1.4995643} {\bibfield  {journal} {\bibinfo  {journal} {J. Chem.
  Phys.}\ }\textbf {\bibinfo {volume} {147}},\ \bibinfo {pages} {224106}
  (\bibinfo {year} {2017})}\BibitemShut {NoStop}%
\bibitem [{\citenamefont {Bennett}\ \emph {et~al.}(2018)\citenamefont
  {Bennett}, \citenamefont {Wang}, \citenamefont {Annaberdiyev}, \citenamefont
  {Melton}, \citenamefont {Shulenburger},\ and\ \citenamefont
  {Mitas}}]{2018BEN}%
  \BibitemOpen
  \bibfield  {author} {\bibinfo {author} {\bibfnamefont {M.~C.}\ \bibnamefont
  {Bennett}}, \bibinfo {author} {\bibfnamefont {G.}~\bibnamefont {Wang}},
  \bibinfo {author} {\bibfnamefont {A.}~\bibnamefont {Annaberdiyev}}, \bibinfo
  {author} {\bibfnamefont {C.~A.}\ \bibnamefont {Melton}}, \bibinfo {author}
  {\bibfnamefont {L.}~\bibnamefont {Shulenburger}}, \ and\ \bibinfo {author}
  {\bibfnamefont {L.}~\bibnamefont {Mitas}},\ }\href {\doibase
  10.1063/1.5038135} {\bibfield  {journal} {\bibinfo  {journal} {J. Chem.
  Phys.}\ }\textbf {\bibinfo {volume} {149}},\ \bibinfo {pages} {104108}
  (\bibinfo {year} {2018})}\BibitemShut {NoStop}%
\bibitem [{\citenamefont {Annaberdiyev}\ \emph {et~al.}(2018)\citenamefont
  {Annaberdiyev}, \citenamefont {Wang}, \citenamefont {Melton}, \citenamefont
  {{Chandler Bennett}}, \citenamefont {Shulenburger},\ and\ \citenamefont
  {Mitas}}]{2018ANN}%
  \BibitemOpen
  \bibfield  {author} {\bibinfo {author} {\bibfnamefont {A.}~\bibnamefont
  {Annaberdiyev}}, \bibinfo {author} {\bibfnamefont {G.}~\bibnamefont {Wang}},
  \bibinfo {author} {\bibfnamefont {C.~A.}\ \bibnamefont {Melton}}, \bibinfo
  {author} {\bibfnamefont {M.}~\bibnamefont {{Chandler Bennett}}}, \bibinfo
  {author} {\bibfnamefont {L.}~\bibnamefont {Shulenburger}}, \ and\ \bibinfo
  {author} {\bibfnamefont {L.}~\bibnamefont {Mitas}},\ }\href {\doibase
  10.1063/1.5040472} {\bibfield  {journal} {\bibinfo  {journal} {J. Chem.
  Phys.}\ }\textbf {\bibinfo {volume} {149}},\ \bibinfo {pages} {134108}
  (\bibinfo {year} {2018})}\BibitemShut {NoStop}%
\bibitem [{\citenamefont {Burkatzki}\ \emph {et~al.}(2007)\citenamefont
  {Burkatzki}, \citenamefont {Filippi},\ and\ \citenamefont {Dolg}}]{2007BUR}%
  \BibitemOpen
  \bibfield  {author} {\bibinfo {author} {\bibfnamefont {M.}~\bibnamefont
  {Burkatzki}}, \bibinfo {author} {\bibfnamefont {C.}~\bibnamefont {Filippi}},
  \ and\ \bibinfo {author} {\bibfnamefont {M.}~\bibnamefont {Dolg}},\ }\href
  {\doibase 10.1063/1.2741534} {\bibfield  {journal} {\bibinfo  {journal} {J.
  Chem. Phys.}\ }\textbf {\bibinfo {volume} {126}},\ \bibinfo {pages} {234105}
  (\bibinfo {year} {2007})}\BibitemShut {NoStop}%
\bibitem [{\citenamefont {Burkatzki}\ \emph {et~al.}(2008)\citenamefont
  {Burkatzki}, \citenamefont {Filippi},\ and\ \citenamefont {Dolg}}]{2008BUR}%
  \BibitemOpen
  \bibfield  {author} {\bibinfo {author} {\bibfnamefont {M.}~\bibnamefont
  {Burkatzki}}, \bibinfo {author} {\bibfnamefont {C.}~\bibnamefont {Filippi}},
  \ and\ \bibinfo {author} {\bibfnamefont {M.}~\bibnamefont {Dolg}},\ }\href
  {\doibase 10.1063/1.2987872} {\bibfield  {journal} {\bibinfo  {journal} {J.
  Chem. Phys.}\ }\textbf {\bibinfo {volume} {129}},\ \bibinfo {pages} {164115}
  (\bibinfo {year} {2008})}\BibitemShut {NoStop}%
\bibitem [{\citenamefont {Umrigar}(1993)}]{1993UMR}%
  \BibitemOpen
  \bibfield  {author} {\bibinfo {author} {\bibfnamefont {C.~J.}\ \bibnamefont
  {Umrigar}},\ }\href@noop {} {\bibfield  {journal} {\bibinfo  {journal} {Phys.
  Rev. Lett.}\ }\textbf {\bibinfo {volume} {71}},\ \bibinfo {pages} {408}
  (\bibinfo {year} {1993})}\BibitemShut {NoStop}%
\bibitem [{\citenamefont {Stedman}\ \emph {et~al.}(1998)\citenamefont
  {Stedman}, \citenamefont {Foulkes},\ and\ \citenamefont {Nekovee}}]{1998STE}%
  \BibitemOpen
  \bibfield  {author} {\bibinfo {author} {\bibfnamefont {M.}~\bibnamefont
  {Stedman}}, \bibinfo {author} {\bibfnamefont {W.}~\bibnamefont {Foulkes}}, \
  and\ \bibinfo {author} {\bibfnamefont {M.}~\bibnamefont {Nekovee}},\
  }\href@noop {} {\bibfield  {journal} {\bibinfo  {journal} {J. Chem. Phys.}\
  }\textbf {\bibinfo {volume} {109}},\ \bibinfo {pages} {2630} (\bibinfo {year}
  {1998})}\BibitemShut {NoStop}%
\bibitem [{\citenamefont {Umrigar}\ \emph {et~al.}(1993)\citenamefont
  {Umrigar}, \citenamefont {Nightingale},\ and\ \citenamefont
  {Runge}}]{1993UMR2}%
  \BibitemOpen
  \bibfield  {author} {\bibinfo {author} {\bibfnamefont {C.}~\bibnamefont
  {Umrigar}}, \bibinfo {author} {\bibfnamefont {M.}~\bibnamefont
  {Nightingale}}, \ and\ \bibinfo {author} {\bibfnamefont {K.}~\bibnamefont
  {Runge}},\ }\href@noop {} {\bibfield  {journal} {\bibinfo  {journal} {J.
  Chem. Phys.}\ }\textbf {\bibinfo {volume} {99}},\ \bibinfo {pages} {2865}
  (\bibinfo {year} {1993})}\BibitemShut {NoStop}%
\bibitem [{\citenamefont {Casula}\ \emph {et~al.}(2005)\citenamefont {Casula},
  \citenamefont {Filippi},\ and\ \citenamefont {Sorella}}]{2005CAS}%
  \BibitemOpen
  \bibfield  {author} {\bibinfo {author} {\bibfnamefont {M.}~\bibnamefont
  {Casula}}, \bibinfo {author} {\bibfnamefont {C.}~\bibnamefont {Filippi}}, \
  and\ \bibinfo {author} {\bibfnamefont {S.}~\bibnamefont {Sorella}},\ }\href
  {\doibase 10.1103/PhysRevLett.95.100201} {\bibfield  {journal} {\bibinfo
  {journal} {Phys. Rev. Lett.}\ }\textbf {\bibinfo {volume} {95}},\ \bibinfo
  {pages} {100201} (\bibinfo {year} {2005})}\BibitemShut {NoStop}%
\bibitem [{\citenamefont {Nakano}\ \emph {et~al.}(2019)\citenamefont {Nakano},
  \citenamefont {Maezono},\ and\ \citenamefont {Sorella}}]{2019NAK}%
  \BibitemOpen
  \bibfield  {author} {\bibinfo {author} {\bibfnamefont {K.}~\bibnamefont
  {Nakano}}, \bibinfo {author} {\bibfnamefont {R.}~\bibnamefont {Maezono}}, \
  and\ \bibinfo {author} {\bibfnamefont {S.}~\bibnamefont {Sorella}},\ }\href
  {\doibase 10.1021/acs.jctc.9b00295} {\bibfield  {journal} {\bibinfo
  {journal} {J. Chem. Theory Comput.}\ }\textbf {\bibinfo {volume} {15}},\
  \bibinfo {pages} {4044} (\bibinfo {year} {2019})}\BibitemShut {NoStop}%
\bibitem [{Note1()}]{Note1}%
  \BibitemOpen
  \bibinfo {note} {The computational cost discussed here is not an actual CPU
  time but an acceptance ratio of off-diagonal trial moves in the
  Metropolis-Hastings algorithm. An actual CPU time is discussed
  later.}\BibitemShut {Stop}%
\bibitem [{\citenamefont {Landau}\ and\ \citenamefont
  {Lifshitz}(1958)}]{1958LAN}%
  \BibitemOpen
  \bibfield  {author} {\bibinfo {author} {\bibfnamefont {L.~D.}\ \bibnamefont
  {Landau}}\ and\ \bibinfo {author} {\bibfnamefont {E.}~\bibnamefont
  {Lifshitz}},\ }\href@noop {} {\emph {\bibinfo {title} {{Quantum Mechanics :
  Non-Relativistic Theory.}}}}\ (\bibinfo  {publisher} {Pergamon Press},\
  \bibinfo {address} {London},\ \bibinfo {year} {1958})\BibitemShut {NoStop}%
\bibitem [{\citenamefont {Slater}(1930)}]{1930SLA}%
  \BibitemOpen
  \bibfield  {author} {\bibinfo {author} {\bibfnamefont {J.~C.}\ \bibnamefont
  {Slater}},\ }\href {\doibase 10.1103/PhysRev.36.57} {\bibfield  {journal}
  {\bibinfo  {journal} {Phys. Rev.}\ }\textbf {\bibinfo {volume} {36}},\
  \bibinfo {pages} {57} (\bibinfo {year} {1930})}\BibitemShut {NoStop}%
\bibitem [{\citenamefont {Clementi}\ and\ \citenamefont
  {Raimondi}(1963)}]{1963CLE}%
  \BibitemOpen
  \bibfield  {author} {\bibinfo {author} {\bibfnamefont {E.}~\bibnamefont
  {Clementi}}\ and\ \bibinfo {author} {\bibfnamefont {D.~L.}\ \bibnamefont
  {Raimondi}},\ }\href {\doibase 10.1063/1.1733573} {\bibfield  {journal}
  {\bibinfo  {journal} {J. Chem. Phys.}\ }\textbf {\bibinfo {volume} {38}},\
  \bibinfo {pages} {2657} (\bibinfo {year} {1963})}\BibitemShut {NoStop}%
\bibitem [{\citenamefont {Clementi}\ \emph {et~al.}(1967)\citenamefont
  {Clementi}, \citenamefont {Raimondi},\ and\ \citenamefont
  {Reinhardt}}]{1967CLE}%
  \BibitemOpen
  \bibfield  {author} {\bibinfo {author} {\bibfnamefont {E.}~\bibnamefont
  {Clementi}}, \bibinfo {author} {\bibfnamefont {D.~L.}\ \bibnamefont
  {Raimondi}}, \ and\ \bibinfo {author} {\bibfnamefont {W.~P.}\ \bibnamefont
  {Reinhardt}},\ }\href {\doibase 10.1063/1.1712084} {\bibfield  {journal}
  {\bibinfo  {journal} {J. Chem. Phys.}\ }\textbf {\bibinfo {volume} {47}},\
  \bibinfo {pages} {5648} (\bibinfo {year} {1967})}\BibitemShut {NoStop}%
\bibitem [{\citenamefont {Ceperley}(1986)}]{1986CEP}%
  \BibitemOpen
  \bibfield  {author} {\bibinfo {author} {\bibfnamefont {D.}~\bibnamefont
  {Ceperley}},\ }\href@noop {} {\bibfield  {journal} {\bibinfo  {journal} {J.
  Stat. Phys.}\ }\textbf {\bibinfo {volume} {43}},\ \bibinfo {pages} {815}
  (\bibinfo {year} {1986})}\BibitemShut {NoStop}%
\bibitem [{\citenamefont {Hammond}\ \emph {et~al.}(1987)\citenamefont
  {Hammond}, \citenamefont {Reynolds},\ and\ \citenamefont
  {Lester~Jr}}]{1987HAM}%
  \BibitemOpen
  \bibfield  {author} {\bibinfo {author} {\bibfnamefont {B.~L.}\ \bibnamefont
  {Hammond}}, \bibinfo {author} {\bibfnamefont {P.~J.}\ \bibnamefont
  {Reynolds}}, \ and\ \bibinfo {author} {\bibfnamefont {W.~A.}\ \bibnamefont
  {Lester~Jr}},\ }\href@noop {} {\bibfield  {journal} {\bibinfo  {journal} {J.
  Chem. Phys.}\ }\textbf {\bibinfo {volume} {87}},\ \bibinfo {pages} {1130}
  (\bibinfo {year} {1987})}\BibitemShut {NoStop}%
\bibitem [{\citenamefont {Ma}\ \emph {et~al.}(2005)\citenamefont {Ma},
  \citenamefont {Drummond}, \citenamefont {Towler},\ and\ \citenamefont
  {Needs}}]{2005MA}%
  \BibitemOpen
  \bibfield  {author} {\bibinfo {author} {\bibfnamefont {A.}~\bibnamefont
  {Ma}}, \bibinfo {author} {\bibfnamefont {N.~D.}\ \bibnamefont {Drummond}},
  \bibinfo {author} {\bibfnamefont {M.~D.}\ \bibnamefont {Towler}}, \ and\
  \bibinfo {author} {\bibfnamefont {R.~J.}\ \bibnamefont {Needs}},\ }\href
  {\doibase 10.1103/PhysRevE.71.066704} {\bibfield  {journal} {\bibinfo
  {journal} {Phys. Rev. E}\ }\textbf {\bibinfo {volume} {71}},\ \bibinfo
  {pages} {066704} (\bibinfo {year} {2005})}\BibitemShut {NoStop}%
\bibitem [{Note2()}]{Note2}%
  \BibitemOpen
  \bibinfo {note} {The calculations were performed on 8 Intel Xeon
  E5-2680v2-2.8~GHTz CPUs ({\protect \it {i.e.}}, 320 cores) installed on a SGI
  cluster.}\BibitemShut {Stop}%
\bibitem [{Note3()}]{Note3}%
  \BibitemOpen
  \bibinfo {note} {The accelerations of actual CPU times are a slightly smaller
  than those of acceptance ratios ({\protect \it {e.g.}}, $\times $ 3.9 and
  $\times $ 2.3 for CPU time and acceptance ratio, respectively, in the neon
  atom). This is because the double-grid algorithm consumes more CPU times when
  computing the discretized laplacians and potentials.}\BibitemShut {Stop}%
\bibitem [{Note4()}]{Note4}%
  \BibitemOpen
  \bibinfo {note} {For the benzene molecule, we measured the CPU times for a
  fixed reference error in the total energy at $a$ = 0.1 Bohr and $a$ = 0.3
  Bohr for the all-electron and the pseudopotential calculations, respectively.
  We determined these values ($a_{\protect \rm {min}}$) such that the
  extrapolation error obtained with 3 parameter polynomial fit ($E\left ( a
  \right ) = {E_0} + {k_1} \cdot {a^2} + {k_2} \cdot {a^4}$) of independent
  energy calculations corresponding to 8 different values of $a \ge a_{\protect
  \rm {min}}$ becomes $\sim $ 2.0 mHa referenced to the safest extrapolation
  value ({\protect \it {i.e.}}, the smallest $a_{\protect \rm {min}}$). Notice
  that the energy-consistent BFD pseudopotentials with the VDZ basis were
  employed.}\BibitemShut {Stop}%
\end{thebibliography}%


\begin{thebibliography}{16}%
\makeatletter
\providecommand \@ifxundefined [1]{%
 \@ifx{#1\undefined}
}%
\providecommand \@ifnum [1]{%
 \ifnum #1\expandafter \@firstoftwo
 \else \expandafter \@secondoftwo
 \fi
}%
\providecommand \@ifx [1]{%
 \ifx #1\expandafter \@firstoftwo
 \else \expandafter \@secondoftwo
 \fi
}%
\providecommand \natexlab [1]{#1}%
\providecommand \enquote  [1]{``#1''}%
\providecommand \bibnamefont  [1]{#1}%
\providecommand \bibfnamefont [1]{#1}%
\providecommand \citenamefont [1]{#1}%
\providecommand \href@noop [0]{\@secondoftwo}%
\providecommand \href [0]{\begingroup \@sanitize@url \@href}%
\providecommand \@href[1]{\@@startlink{#1}\@@href}%
\providecommand \@@href[1]{\endgroup#1\@@endlink}%
\providecommand \@sanitize@url [0]{\catcode `\\12\catcode `\$12\catcode
  `\&12\catcode `\#12\catcode `\^12\catcode `\_12\catcode `\%12\relax}%
\providecommand \@@startlink[1]{}%
\providecommand \@@endlink[0]{}%
\providecommand \url  [0]{\begingroup\@sanitize@url \@url }%
\providecommand \@url [1]{\endgroup\@href {#1}{\urlprefix }}%
\providecommand \urlprefix  [0]{URL }%
\providecommand \Eprint [0]{\href }%
\providecommand \doibase [0]{http://dx.doi.org/}%
\providecommand \selectlanguage [0]{\@gobble}%
\providecommand \bibinfo  [0]{\@secondoftwo}%
\providecommand \bibfield  [0]{\@secondoftwo}%
\providecommand \translation [1]{[#1]}%
\providecommand \BibitemOpen [0]{}%
\providecommand \bibitemStop [0]{}%
\providecommand \bibitemNoStop [0]{.\EOS\space}%
\providecommand \EOS [0]{\spacefactor3000\relax}%
\providecommand \BibitemShut  [1]{\csname bibitem#1\endcsname}%
\let\auto@bib@innerbib\@empty
\bibitem [{\citenamefont {Landau}\ and\ \citenamefont
  {Lifshitz}(1958)}]{1958LAN}%
  \BibitemOpen
  \bibfield  {author} {\bibinfo {author} {\bibfnamefont {L.~D.}\ \bibnamefont
  {Landau}}\ and\ \bibinfo {author} {\bibfnamefont {E.}~\bibnamefont
  {Lifshitz}},\ }\href@noop {} {\emph {\bibinfo {title} {{Quantum Mechanics :
  Non-Relativistic Theory.}}}}\ (\bibinfo  {publisher} {Pergamon Press},\
  \bibinfo {address} {London},\ \bibinfo {year} {1958})\BibitemShut {NoStop}%
\bibitem [{\citenamefont {Chipman}\ and\ \citenamefont
  {Jennings}(1963)}]{1963CHI}%
  \BibitemOpen
  \bibfield  {author} {\bibinfo {author} {\bibfnamefont {D.~R.}\ \bibnamefont
  {Chipman}}\ and\ \bibinfo {author} {\bibfnamefont {L.~D.}\ \bibnamefont
  {Jennings}},\ }\href {\doibase 10.1103/PhysRev.132.728} {\bibfield  {journal}
  {\bibinfo  {journal} {Phys. Rev.}\ }\textbf {\bibinfo {volume} {132}},\
  \bibinfo {pages} {728} (\bibinfo {year} {1963})}\BibitemShut {NoStop}%
\bibitem [{\citenamefont {Slater}(1930)}]{1930SLA}%
  \BibitemOpen
  \bibfield  {author} {\bibinfo {author} {\bibfnamefont {J.~C.}\ \bibnamefont
  {Slater}},\ }\href {\doibase 10.1103/PhysRev.36.57} {\bibfield  {journal}
  {\bibinfo  {journal} {Phys. Rev.}\ }\textbf {\bibinfo {volume} {36}},\
  \bibinfo {pages} {57} (\bibinfo {year} {1930})}\BibitemShut {NoStop}%
\bibitem [{\citenamefont {Clementi}\ and\ \citenamefont
  {Raimondi}(1963)}]{1963CLE}%
  \BibitemOpen
  \bibfield  {author} {\bibinfo {author} {\bibfnamefont {E.}~\bibnamefont
  {Clementi}}\ and\ \bibinfo {author} {\bibfnamefont {D.~L.}\ \bibnamefont
  {Raimondi}},\ }\href {\doibase 10.1063/1.1733573} {\bibfield  {journal}
  {\bibinfo  {journal} {J. Chem. Phys.}\ }\textbf {\bibinfo {volume} {38}},\
  \bibinfo {pages} {2657} (\bibinfo {year} {1963})}\BibitemShut {NoStop}%
\bibitem [{\citenamefont {Clementi}\ \emph {et~al.}(1967)\citenamefont
  {Clementi}, \citenamefont {Raimondi},\ and\ \citenamefont
  {Reinhardt}}]{1967CLE}%
  \BibitemOpen
  \bibfield  {author} {\bibinfo {author} {\bibfnamefont {E.}~\bibnamefont
  {Clementi}}, \bibinfo {author} {\bibfnamefont {D.~L.}\ \bibnamefont
  {Raimondi}}, \ and\ \bibinfo {author} {\bibfnamefont {W.~P.}\ \bibnamefont
  {Reinhardt}},\ }\href {\doibase 10.1063/1.1712084} {\bibfield  {journal}
  {\bibinfo  {journal} {J. Chem. Phys.}\ }\textbf {\bibinfo {volume} {47}},\
  \bibinfo {pages} {5648} (\bibinfo {year} {1967})}\BibitemShut {NoStop}%
\bibitem [{\citenamefont {Sorella}()}]{2019SOR}%
  \BibitemOpen
  \bibfield  {author} {\bibinfo {author} {\bibfnamefont {S.}~\bibnamefont
  {Sorella}},\ }\href@noop {} {\enquote {\bibinfo {title} {{TurboRVB:Quantum
  Monte Carlo Software for Electronic Structure Calculations}},}\ }\bibinfo
  {howpublished} {\url{https://people.sissa.it/~sorella/web}},\ \bibinfo {note}
  {[Online; accessed 30-August-2019]}\BibitemShut {NoStop}%
\bibitem [{\citenamefont {Casula}\ and\ \citenamefont
  {Sorella}(2003)}]{2003CAS}%
  \BibitemOpen
  \bibfield  {author} {\bibinfo {author} {\bibfnamefont {M.}~\bibnamefont
  {Casula}}\ and\ \bibinfo {author} {\bibfnamefont {S.}~\bibnamefont
  {Sorella}},\ }\href@noop {} {\bibfield  {journal} {\bibinfo  {journal} {J.
  Chem. Phys.}\ }\textbf {\bibinfo {volume} {119}},\ \bibinfo {pages} {6500}
  (\bibinfo {year} {2003})}\BibitemShut {NoStop}%
\bibitem [{\citenamefont {Feller}(1996)}]{1996FEL}%
  \BibitemOpen
  \bibfield  {author} {\bibinfo {author} {\bibfnamefont {D.}~\bibnamefont
  {Feller}},\ }\href {\doibase
  10.1002/(SICI)1096-987X(199610)17:13<1571::AID-JCC9>3.0.CO;2-P} {\bibfield
  {journal} {\bibinfo  {journal} {J. Comput. Chem.}\ }\textbf {\bibinfo
  {volume} {17}},\ \bibinfo {pages} {1571} (\bibinfo {year}
  {1996})}\BibitemShut {NoStop}%
\bibitem [{\citenamefont {Schuchardt}\ \emph {et~al.}(2007)\citenamefont
  {Schuchardt}, \citenamefont {Didier}, \citenamefont {Elsethagen},
  \citenamefont {Sun}, \citenamefont {Gurumoorthi}, \citenamefont {Chase},
  \citenamefont {Li},\ and\ \citenamefont {Windus}}]{2007SCU}%
  \BibitemOpen
  \bibfield  {author} {\bibinfo {author} {\bibfnamefont {K.~L.}\ \bibnamefont
  {Schuchardt}}, \bibinfo {author} {\bibfnamefont {B.~T.}\ \bibnamefont
  {Didier}}, \bibinfo {author} {\bibfnamefont {T.}~\bibnamefont {Elsethagen}},
  \bibinfo {author} {\bibfnamefont {L.}~\bibnamefont {Sun}}, \bibinfo {author}
  {\bibfnamefont {V.}~\bibnamefont {Gurumoorthi}}, \bibinfo {author}
  {\bibfnamefont {J.}~\bibnamefont {Chase}}, \bibinfo {author} {\bibfnamefont
  {J.}~\bibnamefont {Li}}, \ and\ \bibinfo {author} {\bibfnamefont {T.~L.}\
  \bibnamefont {Windus}},\ }\href {\doibase 10.1021/CI600510J} {\bibfield
  {journal} {\bibinfo  {journal} {J. Chem. Inf. Model.}\ }\textbf {\bibinfo
  {volume} {47}},\ \bibinfo {pages} {1045} (\bibinfo {year}
  {2007})}\BibitemShut {NoStop}%
\bibitem [{\citenamefont {Sorella}\ \emph {et~al.}(2007)\citenamefont
  {Sorella}, \citenamefont {Casula},\ and\ \citenamefont {Rocca}}]{2007SOR}%
  \BibitemOpen
  \bibfield  {author} {\bibinfo {author} {\bibfnamefont {S.}~\bibnamefont
  {Sorella}}, \bibinfo {author} {\bibfnamefont {M.}~\bibnamefont {Casula}}, \
  and\ \bibinfo {author} {\bibfnamefont {D.}~\bibnamefont {Rocca}},\
  }\href@noop {} {\bibfield  {journal} {\bibinfo  {journal} {J. Chem. Phys.}\
  }\textbf {\bibinfo {volume} {127}},\ \bibinfo {pages} {014105} (\bibinfo
  {year} {2007})}\BibitemShut {NoStop}%
\bibitem [{\citenamefont {Umrigar}\ \emph {et~al.}(2007)\citenamefont
  {Umrigar}, \citenamefont {Toulouse}, \citenamefont {Filippi}, \citenamefont
  {Sorella},\ and\ \citenamefont {Rhenning}}]{2007UMR}%
  \BibitemOpen
  \bibfield  {author} {\bibinfo {author} {\bibfnamefont {C.~J.}\ \bibnamefont
  {Umrigar}}, \bibinfo {author} {\bibfnamefont {J.}~\bibnamefont {Toulouse}},
  \bibinfo {author} {\bibfnamefont {C.}~\bibnamefont {Filippi}}, \bibinfo
  {author} {\bibfnamefont {S.}~\bibnamefont {Sorella}}, \ and\ \bibinfo
  {author} {\bibfnamefont {H.}~\bibnamefont {Rhenning}},\ }\href@noop {}
  {\bibfield  {journal} {\bibinfo  {journal} {Phys. Rev. Lett.}\ }\textbf
  {\bibinfo {volume} {98}},\ \bibinfo {pages} {110201} (\bibinfo {year}
  {2007})}\BibitemShut {NoStop}%
\bibitem [{\citenamefont {Ma}\ \emph {et~al.}(2005)\citenamefont {Ma},
  \citenamefont {Drummond}, \citenamefont {Towler},\ and\ \citenamefont
  {Needs}}]{2005MA}%
  \BibitemOpen
  \bibfield  {author} {\bibinfo {author} {\bibfnamefont {A.}~\bibnamefont
  {Ma}}, \bibinfo {author} {\bibfnamefont {N.~D.}\ \bibnamefont {Drummond}},
  \bibinfo {author} {\bibfnamefont {M.~D.}\ \bibnamefont {Towler}}, \ and\
  \bibinfo {author} {\bibfnamefont {R.~J.}\ \bibnamefont {Needs}},\ }\href
  {\doibase 10.1103/PhysRevE.71.066704} {\bibfield  {journal} {\bibinfo
  {journal} {Phys. Rev. E}\ }\textbf {\bibinfo {volume} {71}},\ \bibinfo
  {pages} {066704} (\bibinfo {year} {2005})}\BibitemShut {NoStop}%
\bibitem [{\citenamefont {Pekeris}(1958)}]{1958PEK}%
  \BibitemOpen
  \bibfield  {author} {\bibinfo {author} {\bibfnamefont {C.~L.}\ \bibnamefont
  {Pekeris}},\ }\href
  {https://journals.aps.org/pr/pdf/10.1103/PhysRev.112.1649} {\bibfield
  {journal} {\bibinfo  {journal} {Phys. Rev.}\ }\textbf {\bibinfo {volume}
  {112}},\ \bibinfo {pages} {1649} (\bibinfo {year} {1958})}\BibitemShut
  {NoStop}%
\bibitem [{\citenamefont {Davidson}\ \emph {et~al.}(1991)\citenamefont
  {Davidson}, \citenamefont {Hagstrom}, \citenamefont {Chakravorty},
  \citenamefont {{Meiser Umar}},\ and\ \citenamefont {Fischer}}]{1991DAV}%
  \BibitemOpen
  \bibfield  {author} {\bibinfo {author} {\bibfnamefont {E.~R.}\ \bibnamefont
  {Davidson}}, \bibinfo {author} {\bibfnamefont {S.~A.}\ \bibnamefont
  {Hagstrom}}, \bibinfo {author} {\bibfnamefont {S.~J.}\ \bibnamefont
  {Chakravorty}}, \bibinfo {author} {\bibfnamefont {V.}~\bibnamefont {{Meiser
  Umar}}}, \ and\ \bibinfo {author} {\bibfnamefont {C.~F.}\ \bibnamefont
  {Fischer}},\ }\href
  {https://journals.aps.org/pra/pdf/10.1103/PhysRevA.44.7071} {\bibfield
  {journal} {\bibinfo  {journal} {Phys. Rev. A}\ }\textbf {\bibinfo {volume}
  {44}},\ \bibinfo {pages} {7071} (\bibinfo {year} {1991})}\BibitemShut
  {NoStop}%
\bibitem [{\citenamefont {Chakravorty}\ and\ \citenamefont
  {Davidson}(1996)}]{1996CHA}%
  \BibitemOpen
  \bibfield  {author} {\bibinfo {author} {\bibfnamefont {S.~J.}\ \bibnamefont
  {Chakravorty}}\ and\ \bibinfo {author} {\bibfnamefont {E.~R.}\ \bibnamefont
  {Davidson}},\ }\href {https://pubs.acs.org/sharingguidelines} {\bibfield
  {journal} {\bibinfo  {journal} {J. Phys. Chem.}\ }\textbf {\bibinfo {volume}
  {100}},\ \bibinfo {pages} {6167} (\bibinfo {year} {1996})}\BibitemShut
  {NoStop}%
\bibitem [{\citenamefont {Clementi}\ and\ \citenamefont
  {Hofmann}(1995)}]{1995CLE}%
  \BibitemOpen
  \bibfield  {author} {\bibinfo {author} {\bibfnamefont {E.}~\bibnamefont
  {Clementi}}\ and\ \bibinfo {author} {\bibfnamefont {D.}~\bibnamefont
  {Hofmann}},\ }\href {\doibase 10.1016/0166-1280(94)03814-2} {\bibfield
  {journal} {\bibinfo  {journal} {J. Mol. Struct. THEOCHEM}\ }\textbf {\bibinfo
  {volume} {330}},\ \bibinfo {pages} {17} (\bibinfo {year} {1995})}\BibitemShut
  {NoStop}%
\end{thebibliography}%

\end{document}


\title{Supplemental material for Speeding up the {\it ab initio} diffusion Monte Carlo by a smart lattice regularization}

\author{Kousuke Nakano$^{1}$}
\email{kousuke\_1123@icloud.com}

\author{Ryo Maezono$^{2,3}$}

\author{Sandro Sorella$^{1}$}
\email{sorella@sissa.it}
  
\affiliation{$^{1}$ International School for Advanced Studies (SISSA),
  Via Bonomea 265, 34136, Trieste, Italy}

\affiliation{$^{2}$ School of Information Science, Japan Advanced Institute of Science and Technology (JAIST), Asahidai 1-1, Nomi, Ishikawa 923-1292, Japan}

\affiliation{$^{3}$ Computational Engineering Applications Unit, RIKEN, 2-1 Hirosawa, Wako, Saitama 351-0198, Japan}

\date{\today}

\begin{abstract}
This supplemental material gives details about the scaling of the number of electrons within $r_c$ obtained by the Thomas-Fermi model (Sec.~I), a remedy for smooth extrapolations (Sec.~II), and the variational (VMC) and the lattice regularized quantum Monte Carlo (LRDMC) calculations mentioned in the main text (Sec.~III).


\end{abstract}
\maketitle

\makeatletter
\def\Hline{%
\noalign{\ifnum0=`}\fi\hrule \@height 1pt \futurelet
\reserved@a\@xhline}
\makeatother

\makeatletter
\def\Bline{%
\noalign{\ifnum0=`}\fi\hrule \@height 0.5pt \futurelet
\reserved@a\@xhline}
\makeatother


\makeatletter
\renewcommand{\refname}{}
\renewcommand*{\citenumfont}[1]{S#1}
\renewcommand*{\bibnumfmt}[1]{[S#1]}
\makeatother

\setcounter{table}{0}
\setcounter{equation}{0}
\setcounter{figure}{0}
\renewcommand{\thetable}{S-\Roman{table}}
\renewcommand{\thefigure}{S-\arabic{figure}}
\renewcommand{\theequation}{S-\arabic{equation}}

\section{The scaling of the number of electrons within $r_c$ obtained by the Thomas-Fermi model}

\begin{figure}[htbp]
  \centering
  \includegraphics[width=7.6cm]{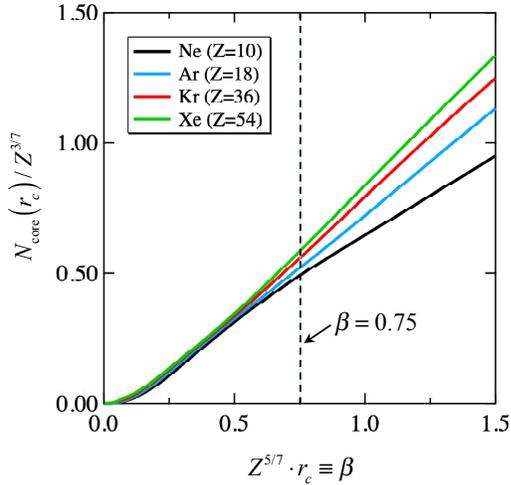}
  \caption{The plot of ${N_{{\text{core}}}}\left( {{r_c}} \right) / Z^{3/7}$ obtained by VMC calculations v.s  $\beta$, showing that the scaling (${N_{\rm{core}}}\left( {{r_c}} \right) \propto {Z^{3/7}}$) is satisfied at small $\beta$ region. $\beta$ = 0.75 is employed in the present work.}
  \label{RDF_scaling}
\end{figure}

\vspace{3mm}
According to the Thomas-Fermi model~{\cite{1958LAN}}, the electron density of the atomic number $Z$ can be represented by:
\begin{equation}
\rho \left( r \right) = {Z^2}f\left( {\frac{{{Z^{1/3}}r}}{b}} \right),
\end{equation}
where $b$ is the constant value $b = {\left( {{{9{\pi ^2}}}/{{128}}} \right)^{1/3}}$, $f\left( x \right)$ is:
\begin{equation}
f\left( x \right) = \frac{{32}}{{9\pi }}{\left( {\frac{{\chi \left( x \right)}}{x}} \right)^{\frac{3}{2}}},
\end{equation}
and ${\chi \left( x \right)}$ is the universal function independent of $Z$.
Therefore, the number of electrons within $r_c$ is defined by:
\begin{equation}
{N_{{\text{core}}}}\left( {{r_c}} \right) = \int_0^\infty  {dr\,\,4\pi {r^2}{Z^2}f\left( {\frac{{{Z^{1/3}}r}}{b}} \right)} \exp \left( { - \frac{r^2}{{2r_c^2}}} \right).
\label{Ncore_TF}
\end{equation}
When ${r}$ is replaced with ${Z^{1/3}}r/b=x$, the number of electrons within $r_c$ is represented as:
\begin{equation}
{\rm{Eq}}.\left( {\ref{Ncore_TF}} \right) \Leftrightarrow Z\int_0^\infty  {dx\,\,{x^{\frac{1}{2}}}\chi {{\left( x \right)}^{\frac{3}{2}}}} \exp \left( { - \xi {x^2}} \right),
\label{eq_RDF_constant}
\end{equation}
where $\xi  = {\left( {b{Z^{ - 1/3}}/\sqrt 2 {r_c}} \right)^2}$.
$\chi \left( x \right)$ can be approximated by the following polynominal expression at small $x$ region ($x \ll 1$)~{\cite{1958LAN}}:
\begin{equation}
\chi \left( x \right) =  1 - Ax +  \cdots
\end{equation}
where $A$ is the constant value $A$ = 1.8858. 
If $\xi$ is large enough ({\it {i.e.}}, $ {Z^{1/3}}{r_c}/b \ll 1$), only small $x$ region contributes to the integral and the high-order terms can be neglected.
Therefore, the above equation can be approximated by:
\begin{equation}
{N_{{\text{core}}}}\left( {{r_c}} \right) \simeq Z\int_0^\infty  {dx\,\,{x^{\frac{1}{2}}}} \exp \left( { - \xi {x^2}} \right).
\end{equation}
Since the integral can be replaced by the gamma function:
\begin{equation}
\int_0^\infty  {dx\,\,{x^{2s - 1}}\exp \left( { - \xi {x^2}} \right)}  = \frac{1}{2}\Gamma \left( s \right){\xi ^{ - s}},
\end{equation}
the number of electrons within $r_c$ can be represented as:
\begin{equation}
{N_{{\text{core}}}}\left( {{r_c}} \right) \simeq \frac{1}{2}Z \cdot \Gamma \left( {\frac{3}{4}} \right){\xi ^{ - \frac{3}{4}}} \equiv \frac{1}{2}{\left( {\frac{b}{{\sqrt 2 }}} \right)^{ - \frac{3}{2}}}\Gamma \left( {\frac{3}{4}} \right){\left( {Z \cdot {r_c}} \right)^{\frac{3}{2}}}.
\end{equation}
By substituting ${r_c}$ with $\beta  \cdot {Z^{ - \theta }}$, we finally get the relation:
\begin{equation}
{N_{{\text{core}}}}\left( {{r_c}} \right) \simeq \frac{1}{2}{\left( {\frac{b}{{\sqrt 2 \beta }}} \right)^{ - \frac{3}{2}}}\Gamma \left( {\frac{3}{4}} \right){Z^{\frac{3}{2}\left( {1 - \theta } \right)}} \propto {Z^{\frac{3}{2}\left( {1 - \theta } \right)}},
\end{equation}
and for $\theta  = 5/7$:
\begin{equation}
{N_{{\text{core}}}}\left( {{r_c}} \right) \propto {Z^{\frac{3}{7}}}.
\label{N_core_appx}
\end{equation}
Eq.~(\ref{N_core_appx}) is valid only when the inequality $\xi \gg 1$ is satisfied. This depends on the prefactor $\beta$ as well as the atomic number $Z$. In practice, the prefactor $\beta$ should be small enough so that ${N_{\rm{core}}}\left( {{r_c}} \right) \propto {Z^{3/7}}$ is valid in a wide range of $Z$ values, even outside the asymptotic power law regime. Fig.{~{\ref{RDF_scaling}}} shows that the plot of ${N_{{\text{core}}}}\left( {{r_c}} \right)$ divided by $Z^{3/7}$ obtained by VMC calculations  v.s  $\beta$ for Ne, Ar, Kr and Xe atoms. This figure shows that $\beta$ = 0.75 is small enough to satisfy the above scaling.

\begin{figure}[htbp]
  \centering
  \includegraphics[width=7.6cm]{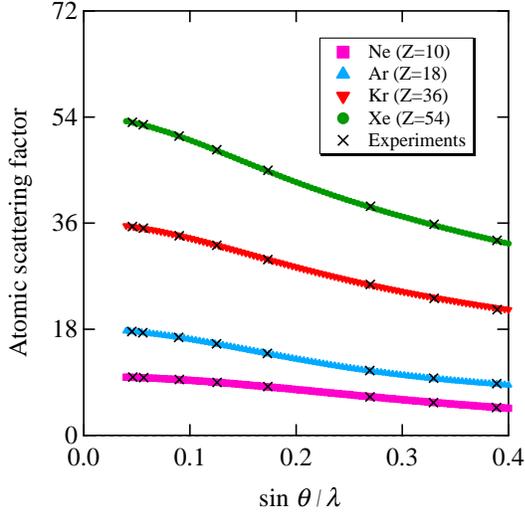}
  \caption{Atomic scattering factors obtained by the VMC calculations, and those obtained by scattering X-ray measurements{~\cite{1963CHI}}.}
  \label{ASF}
\end{figure}

\vspace{3mm}
The electron densities obtained by the VMC calculations were validated by comparison with the experimental atomic scattering factors (ASFs), as shown in Fig.~{\ref{ASF}}.
ASFs can be readily calculated using the electron densities obtained by the {\it {ab initio}} VMC calculations according to the following relation{~\cite{1963CHI}}:
\begin{equation}
{\rm{ASF}} = \int_0^\infty  {4\pi {r^2}\rho \left( r \right)\frac{{\sin kr}}{{kr}}dr},
\end{equation}
where $k = 4\pi \sin \theta /\lambda$, $2\theta$ is the scattering angle and $\lambda$ is the wavelength. TurboRVB enables us to calculate a radial distribution function as well as an electron density from a many-body wave function both in VMC and DMC levels using the forward walking technique.

\section{A remedy for smooth extrapolations}
In order to improve the quality of the energy extrapolation for $a \to 0$, it is important to increase $r_c$ as $a$ increases.
This is because if $r_c$ is fixed ({\it {i.e.}}, $a'/a$ is also fixed according to Eq.~(6) in the main text), $a^\prime \simeq r_c$ is no longer satisfied in a large $a$ region, which introduces a large bias by the larger lattice space $a'$ especially in the vicinity of the border between the core and valence regions. A simple parametrization to solve this problem is ${r_c}\left( {a,Z} \right) = {r_c}\left( Z \right) f\left( a \right)$, where $f\left( a \right)$ is an arbitrary function satisfying $f\left( 0 \right) = 1$ and $f\left( \infty  \right) = const.$ In this study, a simple polynominal function:
\begin{equation}
f\left( a \right) = \frac{{\kappa {{\left( {Z \cdot a} \right)}^2} + 1}}{{{{\left( {Z \cdot a} \right)}^2} + 1}} \equiv \frac{{\kappa \cdot {\alpha ^{ - 2}} + 1}}{{{\alpha ^{ - 2}} + 1}}
\end{equation}
\label{f_a}
is employed, where $\kappa$ is a prefactor, and $a = {\left( {\alpha  \cdot Z} \right)^{ - 1}}$. 
Therefore, ${r_c}\left( {a,Z} \right) $ can be parametrized as:
\begin{equation}
{r_c}\left( {a,Z} \right) = {r_c}\left( Z \right)f\left( a \right) \equiv \frac{{\beta \left( {\kappa  \cdot {\alpha ^{ - 2}} + 1} \right)}}{{{\alpha ^{ - 2}} + 1}} \cdot {Z^{ - \frac{5}{7}}}.
\label{r_c}
\end{equation}
Eq.~(6) in the main text indicates that ${N_{\rm{core}}}\left( {{r_c}} \right)$ should be smaller than ${N_{\rm{valence}}}\left( {{r_c}} \right)$ for any $a$ and $Z$, otherwise the double-grid LRDMC becomes useless ({\it {i.e.}}, $a'/a < 1$) in a certain case. According to the Thomas-Fermi theory, ${N_{\rm{core}}}\left( {{r_c}} \right)$ becomes equal to ${N_{\rm{valence}}}\left( {{r_c}} \right)$ at ${r_c} = 1.33{Z^{ - 1/3}}$~{\cite{1958LAN}}. Therefore, the following inequality should be satisfied for all $Z$ and $a$,
\begin{equation}
{r_c}\left( {a,Z} \right) < 1.33{Z^{ - \frac{1}{3}}}\,\,\;(\forall a \in a > 0,\forall Z \in Z \geqslant 3).
\end{equation}
Thus, we obtain $\kappa < 2.69$ for $\beta$ = 0.75. $\kappa$ = 2.5 is employed here. The new algorithm determines ${r_c}\left( {a,Z} \right)$ using $Z$ and $a$ according to Eq.~({\ref{r_c}}), then, a corresponding proper $a'/a$ is calculated by the obtained ${r_c}\left( {a,Z} \right)$ according to Eq.~(6) in the main text, wherein ${N_{\rm{core}}}\left( {{r_c}} \right)$ and ${N_{\rm{valence}}}\left( {{r_c}} \right)$ are estimated by the Slater's effective models~{\cite{1930SLA}} with the exponents that Clementi proposed based on HF calculations~{\cite{1963CLE, 1967CLE}}. 

\begin{figure*}[htbp]
  \centering
  \includegraphics[width=16.8cm]{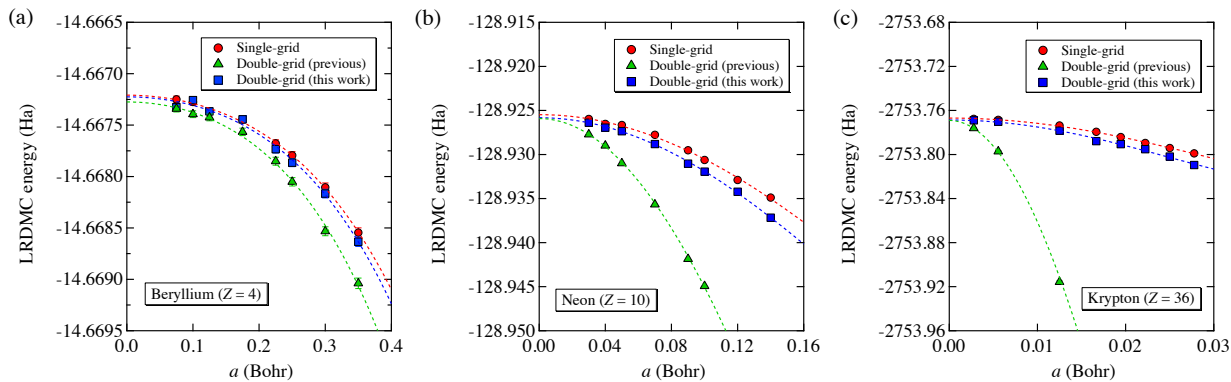}
  \caption{LRDMC energies of (a) Be ($Z$ = 4), (b) Ne ($Z$ = 10) and (c) Kr ($Z$ = 36) atoms obtained by the single-grid (red circles), the previous (green triangles) and the newly developed (blue squares) double-grid algorithms.}
  \label{sup_fig_lrdmc_single_double}
\end{figure*}

\vspace{3mm}
In Fig.~{\ref{sup_fig_lrdmc_single_double}} and Table~{\ref{sup_table_single_double_grid}}, we show the LRDMC energies of Be ($Z$ = 4), Ne ($Z$ = 10) and Kr ($Z$ = 36) atoms obtained by the single-grid, the previous and the new double-grid algorithms with the above remedy.
We remark that in Figure~{\ref{sup_fig_lrdmc_single_double}} the  LRDMC energies obtained by the previous parametrization are significantly biased in the  large $a$ region especially in Kr ($Z$ = 36). 
On the other hand, our new parametrization suppresses these significant biases, and the obtained LRDMC energies in the small $a$ region ($a \to 0$) are essentially unbiased for all $Z$.
In Tables~{\ref{sup_table_single_double_grid}}, it is evident that ${r_c}\left( {a,Z} \right)$ increases as $a$ increases, and $a'/a$ decreases as $a$ increases, by which the condition $a^\prime \simeq r_c$ is satisfied for any $a$ and $Z$. In this way, unnecessary large biases ( i.e. not saving computational time) are suppressed, and smooth extrapolations are achieved. Notice that Table~{\ref{sup_table_single_double_grid}} indicates that this modified algorithm implies smaller speedups as $a$ increases. This is because ${r_c}\left( {a,Z} \right)$ (Eq.~({\ref{r_c}})) becomes slightly larger than the original ${r_c}\left( Z \right)$ due to $f\left( a \right)$ as $a$ increases. However, the effect is negligible in practice because small $a$ calculations are much more important than large ones ({\it {i.e.}}, the computational cost is proportional to $a^{-2}$). Thus, within the remedy, the new double-grid algorithm achieves both acceleration and smooth extrapolation.

\section{The details of VMC and LRDMC calculations}

\vspace{3mm}
The variational (VMC) and lattice regularized quantum Monte Carlo (LRDMC) calculations for He, Be, Ne, Ar, Kr, Xe and C$_6$H$_6$ were performed using TurboRVB{\cite{2019SOR}}.
In the VMC calculations, Jastrow Slater (JSD) and Antisymmetrized Geminal Power (JAGP){~\cite{2003CAS}} ansatz were employed, and the cc-pVDZ (He, Be, Ne, Ar, Kr, and C$_6$H$_6$) or ADZP (Xe) basis set taken from EMSL Basis Set Library{~\cite{1996FEL,2007SCU}} were used in the determinant and the Jastrow parts. The variational JSD and JAGP wave functions were optimized using the stochastic configuration in combination with the linear method~{\cite{2007SOR,2007UMR}}, by which all variational parameters in the Jastrow and the determinant parts including the exponents were optimized. In LRDMC calculations, the wave functions optimized based on JAGP ansatz were used for the guiding functions. 
The obtained VMC and LRDMC energies are shown in Table~{\ref{qmc_summary}}.
Thanks to our careful optimizations, our VMC-JSD energies are lower than the previous results, especially when Z becomes larger. Remarkably, the JAGP ansatz further improves the variational energies. Our JAGP ansatz also improves the LRDMC energies ({\it {i.e.}}, the nodal surfaces) as well as the variational energies. 

\onecolumngrid

\begin{center}
\begin{table*}[htbp]
\caption{\label{sup_table_single_double_grid} LRDMC energies of Be, Ne, and Kr atoms obtained by the single and double-grid schemes.}
\vspace{3mm}
\begin{tabular}{c|c|c|c|c|cc|cc}
\Hline
\multirow{2}{*}{Element} & \multicolumn{2}{c|}{Lattice space} & \multicolumn{1}{c|}{Single grid} & \multicolumn{5}{c}{Double grid (this work)} \\
\cline{2-9}
& $a \equiv \left( {\alpha  \cdot Z} \right)^{ - 1} $ &  $\alpha$ & Energy (Ha) & Energy (Ha) & Bias (mHa){\footnotemark[1]} & Acceleration{\footnotemark[2]} & $a'/a$ & ${r_c}\left( {a,Z} \right)$ \\
\Hline
\multirow{10}{*}{Be ($Z$ = 4)}
 & $a \to 0$ & $\alpha  \to \infty$ & -14.66721(28) & -14.66723(28) & 0.0(0) & - & - & - \\
 & 0.08 & 3.33 & -14.66725(32) & -14.66732(32) & 0.1(0) & $\times$ 2.0 & 1.733 & 0.313 \\
 & 0.10 & 2.50 & -14.66727(32) & -14.66726(32) & 0.0(0) & $\times$ 1.9 & 1.654 & 0.336 \\
 & 0.13 & 2.00 & -14.66736(32) & -14.66737(32) & 0.0(0) & $\times$ 1.8 & 1.580 & 0.362 \\
 & 0.18 & 1.43 & -14.66745(34) & -14.66744(34) & 0.0(0) & $\times$ 1.6 & 1.458 & 0.416 \\
 & 0.23 & 1.11 & -14.66768(36) & -14.66773(38) & 0.1(1) & $\times$ 1.5 & 1.372 & 0.466 \\
 & 0.25 & 1.00 & -14.66770(38) & -14.66787(38) & 0.1(1) & $\times$ 1.4 & 1.340 & 0.488 \\
 & 0.30 & 0.83 & -14.66811(42) & -14.66817(42) & 0.1(1) & $\times$ 1.4 & 1.291 & 0.525 \\
 & 0.35 & 0.71 & -14.66855(46) & -14.66863(48) & 0.1(1) & $\times$ 1.3 & 1.257 & 0.555 \\
\hline\Bline \hline\Bline
\multirow{10}{*}{Ne ($Z$ = 10)}
 & $a \to 0$ & $\alpha  \to \infty$ & -128.92548(12) & -128.92583(12) & 0.4(2) & - & - & - \\
 & 0.03 & 3.33 & -128.92597(14) & -128.92638(14) & 0.4(2) & $\times$ 3.5 & 2.449 & 0.163 \\
 & 0.04 & 2.50 & -128.92653(14) & -128.92695(14) & 0.4(2) & $\times$ 3.3 & 2.352 & 0.175 \\
 & 0.05 & 2.00 & -128.92665(14) & -128.92736(14) & 0.7(2) & $\times$ 3.1 & 2.255 & 0.188 \\
 & 0.07 & 1.43 & -128.92776(15) & -128.92881(15) & 1.0(2) & $\times$ 2.7 & 2.080 & 0.216 \\
 & 0.09 & 1.11 & -128.92952(15) & -128.93104(16) & 1.5(2) & $\times$ 2.5 & 1.942 & 0.242 \\
 & 0.10 & 1.00 & -128.93063(15) & -128.93194(16) & 1.3(2) & $\times$ 2.4 & 1.887 & 0.253 \\
 & 0.12 & 0.83 & -128.93289(17) & -128.93422(17) & 1.3(2) & $\times$ 2.2 & 1.797 & 0.273 \\
 & 0.14 & 0.71 & -128.93489(19) & -128.93716(21) & 2.3(4) & $\times$ 2.1 & 1.731 & 0.289 \\
\hline\Bline \hline\Bline
\multirow{9}{*}{Kr ($Z$ = 36)}
 & $a \to 0$ & $\alpha  \to \infty$ & -2753.76693(53) & -2753.76860(50) & 1.7(0.7) & - & - & - \\
 & 0.00278 & 10.00 & -2753.76770(76) & -2753.76891(65) & 1.2(1.0) & $\times$ 7.0 & 3.641 & 0.059 \\
 & 0.00556 & 5.00 & -2753.76856(60) & -2753.77025(64) & 1.7(0.9) & $\times$ 6.7 & 3.541 & 0.061 \\
 & 0.01250 & 2.22 & -2753.77369(58) & -2753.77837(86) & 4.7(1.0) & $\times$ 5.5 & 3.159 & 0.073 \\
 & 0.01667 & 1.67 & -2753.77937(75) & -2753.78784(69) & 8.5(1.0) & $\times$ 4.8 & 2.936 & 0.081 \\
 & 0.01944 & 1.43 & -2753.78412(73) & -2753.79063(73) & 6.5(1.0) & $\times$ 4.5 & 2.809 & 0.087 \\
 & 0.02222 & 1.25 & -2753.78951(77) & -2753.79514(73) & 5.6(1.1) & $\times$ 4.2 & 2.701 & 0.092 \\
 & 0.02500 & 1.11 & -2753.79412(82) & -2753.80190(85) & 7.8(1.2) & $\times$ 4.0 & 2.609 & 0.097 \\
 & 0.02778 & 1.00 & -2753.79899(88) & -2753.80951(66) & 10.5(1.1) & $\times$ 3.8 & 2.532 & 0.101 \\
\Hline
\end{tabular}
\footnotetext[1]{The difference in total energy between the single- and double-grid algorithms.}
\footnotetext[2]{The accelerations were not measured by actual CPU times but by acceptance ratios of trial moves in LRDMC.}
\end{table*}
\end{center}

\begin{center}
\begin{table*}[htbp]
\caption{\label{qmc_summary} Total energies of He, Be, Ne, Ar, Kr and Xe. The LRDMC energies shown here were calculated with $a = {\left( {3.5 Z} \right)^{ - 1}}$, which are consistent with the typical scaling of the time step in the standard DMC ($\tau  \propto {Z^{ - 2}}$).}
\vspace{3mm}
\begin{tabular}{c|c|c|c}
\Hline
Element & Method & Total energy (Ha) & Correlation (\%) \\
\Hline
\multirow{7}{*}{He (Z=2){\footnotemark[1]}}
 & HF{\footnotemark[7]}             & -2.86165214   &  0 \\
\cline{2-4}
 & VMC-JSD{\footnotemark[7]}     & -2.903527(9)  &  99.53(2) \\
 & DMC{\footnotemark[7]}      & -2.903719(2)  &  99.99(0) \\
\cline{2-4}
 & VMC-JSD        & -2.903352(79)  &  99.12(2)  \\
 & VMC-JAGP       & -2.903443(70)  &  99.33(2)  \\
 & LRDMC & -2.903732(62)  &  100.02(1) \\
\cline{2-4}
 & Exact{\footnotemark[8]}  & -2.903724     &  100  \\
\Hline
\multirow{6}{*}{Be (Z=4){\footnotemark[2]}}
 & HF{\footnotemark[9]}              & -14.573023   &  0 \\
\cline{2-4}
 & VMC-AGP{\footnotemark[9]}     & -14.66504(4)  &  97.54(0) \\
 & DMC{\footnotemark[9]}      & -14.66726(1) &  99.89(0) \\
\cline{2-4}
 & VMC-JAGP       & -14.665828(42)  &  98.38(0)  \\
 & LRDMC & -14.667247(31)  &  99.88(0) \\
\cline{2-4}
 & Exact{\footnotemark[10]}          & -14.66736   &  100  \\
\Hline
\multirow{7}{*}{Ne (Z=10){\footnotemark[3]}}
 & HF{\footnotemark[7]}             & -128.5470981  &  0 \\
\cline{2-4}
 & VMC-JSD{\footnotemark[7]}       & -128.891(5)   &  88(1) \\
 & DMC{\footnotemark[7]}       & -128.9231(1)  &  95.9(3) \\
\cline{2-4}
 & VMC-JSD        & -128.89803(12)  & 89.5(12)  \\
 & VMC-JAGP       & -128.90354(44)  & 91.0(11)  \\
 & LRDMC          & -128.92626(13)  & 96.7(3)  \\
\cline{2-4}
 & Exact{\footnotemark[11]}          & -128.939  &  100  \\
\Hline
\multirow{7}{*}{Ar (Z=18){\footnotemark[4]}}
 & HF{\footnotemark[7]}             & -526.8175128  &  0 \\
\cline{2-4}
 & VMC-JSD{\footnotemark[7]}         & -527.3817(2)  &  77.02(3) \\
 & DMC{\footnotemark[7]}             & -527.4840(2)  &  90.99(3) \\
\cline{2-4}
 & VMC-JSD        & -527.41937(25)  &  82.17(5)  \\
 & VMC-JAGP       & -527.43164(37)  &  83.84(5)  \\
 & LRDMC          & -527.49542(18)  &  92.55(3) \\
\cline{2-4}
 & Exact{\footnotemark[12]}          & -527.55  &  100  \\
\Hline
\multirow{7}{*}{Kr (Z=36){\footnotemark[5]}}
 & HF{\footnotemark[7]}             & -2752.054977  &  0 \\
\cline{2-4}
 & VMC-JSD{\footnotemark[7]}         & -2753.2436(6)  &  57.28(3) \\
 & DMC{\footnotemark[7]}             & -2753.7427(6)  &  81.34(3) \\
\cline{2-4}
 & VMC-JSD        & -2753.61420(46)  &  75.14(2)  \\
 & VMC-JAGP       & -2753.62841(44)  &  75.83(2) \\
 & LRDMC          & -2753.77151(78)	 &  82.72(4) \\
\cline{2-4}
 & Exact{\footnotemark[13]}          & -2754.13  &  100  \\
\Hline
\multirow{7}{*}{Xe (Z=54){\footnotemark[6]}}
 & HF{\footnotemark[7]}             & -7232.138363  &  0 \\
\cline{2-4}
 & VMC-JSD{\footnotemark[7]}         & -7233.700(2)  &  45.51(6) \\
 & DMC{\footnotemark[7]}             & -7234.785(1)  &  77.12(3) \\
\cline{2-4}
 & VMC-JSD        & -7234.50730(87)  &  69.03(3)  \\
 & VMC-JAGP       & -7234.56494(82)  &  70.71(2)  \\
 & LRDMC          & -7234.8320(13) & 78.50(4) \\
\cline{2-4}
 & Exact{\footnotemark[13]}          & -7235.57  &  100  \\
\Hline
\end{tabular}
\footnotetext[1]{Our modified cc-pVDZ basis is composed of 3$s$1$p$ (Z $\leq$ 5.77) and 2$s$1$p$ (Z $\leq$ 1.275) for the determinant and Jastrow part, respectively.}
\footnotetext[2]{Our modified cc-pVDZ basis is composed of 7$s$4$p$1$d$ (Z $\leq$ 100.5) and 5$s$3$p$1$d$ (Z $\leq$ 9.17) for the determinant and Jastrow part, respectively.}
\footnotetext[3]{Our modified cc-pVDZ basis is composed of 6$s$4$p$1$d$ (Z $\leq$ 173.5) and 3$s$3$p$1$d$ (Z $\leq$ 7.81) for the determinant and Jastrow part, respectively.}
\footnotetext[4]{Our modified cc-pVDZ basis is composed of 8$s$8$p$1$d$ (Z $\leq$ 459.7) and 6$s$6$p$1$d$ (Z $\leq$ 64.69) for the determinant and Jastrow part, respectively.}
\footnotetext[5]{Our modified cc-pVDZ basis is composed of 11$s$11$p$6$d$ (Z $\le$ 6582.01) and 7$s$8$p$4$d$ (Z $\le$ 129.00) for the determinant and Jastrow part, respectively.}
\footnotetext[6]{Our modified ADZP basis is composed of 15$s$16$p$9$d$2$f$ (Z $\le$ 19789.22) and 11$s$12$p$7$d$ (Z $\le$ 335.98) for the determinant and Jastrow part, respectively.}
\footnotetext[7]{See ref.~{\onlinecite{2005MA}}.} 
\footnotetext[8]{See ref.~{\onlinecite{1958PEK}}.}
\footnotetext[9]{See ref.~{\onlinecite{2003CAS}}.}
\footnotetext[10]{See ref.~{\onlinecite{1958PEK}}.}
\footnotetext[11]{See ref.~{\onlinecite{1991DAV}}.}
\footnotetext[12]{See ref.~{\onlinecite{1996CHA}}.}
\footnotetext[13]{See ref.~{\onlinecite{1995CLE}}.}

\end{table*}
\end{center}

\clearpage

\bibliographystyle{apsrev4-1}
\bibliography{./references.bib}